\documentclass[twocolumn,showpacs,preprintnumbers,amsmath,amssymb]{revtex4-1}

\usepackage{epsfig,psfrag}
\usepackage{dcolumn}
\usepackage{bm}
\usepackage{graphicx}
\usepackage{color}

\begin{document}
\title{Fermi liquid approach to the quantum RC circuit: renormalization-group analysis of the Anderson and Coulomb blockade models}

\author{Michele Filippone}
\author{Christophe Mora}
\affiliation{Laboratoire Pierre Aigrain, \'Ecole Normale
  Sup\'erieure, Universit\'e Paris 7 Diderot, 
CNRS; 24 rue Lhomond, 75005 Paris, France} 


\begin{abstract}
We formulate a general approach for studying the low frequency response
of an interacting quantum dot connected to leads in the presence of
oscillating gate voltages. The energy dissipated is characterized
by the charge relaxation resistance which, under the loose assumption
of Fermi liquid behaviour at low energy, is shown to depend only on
static charge susceptibilities. The predictions of the scattering
theory are recovered in the non-interacting limit while 
the effect of interactions is simply to 
replace densities of states by charge susceptibilities in formulas.
In order to substantiate the Fermi liquid picture in the case
of a quantum RC geometry, we apply a 
renormalization-group analysis and derive the low energy Hamiltonian
for two specific models: the Anderson and the Coulomb blockade
models. The Anderson model is shown, using a field theoretical approach
based on Barnes slave-bosons, to map onto the Kondo model. We recover the
well-known expression of the Kondo temperature for the asymmetric 
Anderson model
and compute the charge susceptibility.
The Barnes slave-bosons are extended to the Coulomb blockade
model where the renormalization-group analysis can be carried out
perturbatively up to zero energy. All calculations agree with the Fermi
liquid nature of the low energy fixed point and satisfy the Friedel sum rule.
\end{abstract}

\pacs{71.10.Ay, 73.63.Kv, 72.15.Qm}

\maketitle

\section{Introduction}

The ability to probe and manipulate electrons in real time 
constitutes one of the main challenges of transport in quantum dots. 
This program is spurred by technological progress in guiding and 
processing high-frequency electronic signals.
At low frequency, charge or spin can be  transferred
adiabatically with quantum pumps~\cite{pothier1992,buttiker1994,brouwer1998,
aleiner1998b,leek2005,Splettstoesser2006} and
single electron tunneling events
can be measured by coupling the system to a nearby 
quantum point contact~\cite{schleser2004,vandersypen2004},
the detection bandwidth is however restricted to the kilohertz regime.
These low frequency experiments may control and monitor 
single-charge transfer events but they are not able to capture
the coherent dynamics of charge carriers.

Early experiments at high frequency used a microwave
source (typically above the GHz) 
to irradiate the quantum dot in the presence of a source-drain DC
bias voltage~\cite{Kouwenhoven1994,kogan2004}.
The energy of a photon $\hbar \omega$
may exceed the thermal energy $k_B T$ and photon-assisted
tunneling takes place. Electrons are then able to tunnel across
the quantum dot by emitting or absorbing photons from the 
microwave signal. 
Probing quantum dots at high frequency provides information
on the quantum motion of electrons~\cite{Schoelkopf1998}. 
The microwave part of
the noise power spectrum corresponds to 
 the typical energies (level spacing, charging energy, etc) 
of quantum dots of micrometer size.
In the quantum regime $\hbar \omega >k_B T $, the noise emitted
by a nanoconductor device can be absorbed and measured by
an on-chip quantum detector. Different schemes have been developed
where the quantum detector, located in the vicinity of the source, 
can be a quantum dot~\cite{aguado2000,onac2006, gustavsson2007}, 
a SIS tunnel junction~\cite{deblock2003,billangeon2006,basset2010} 
or a superconducting
resonator~\cite{xue2009}. 
An alternative way to measure directly the emitted noise
from the nanoconductor is to use cryogenic low noise
amplifiers~\cite{zakka2007,gabelli2008,zakka2010}.
This non-exhaustive synopsis of high frequency experiments 
demonstrates the vitality of research in this
field.

\begin{figure}[b]
    \includegraphics[width=8cm]{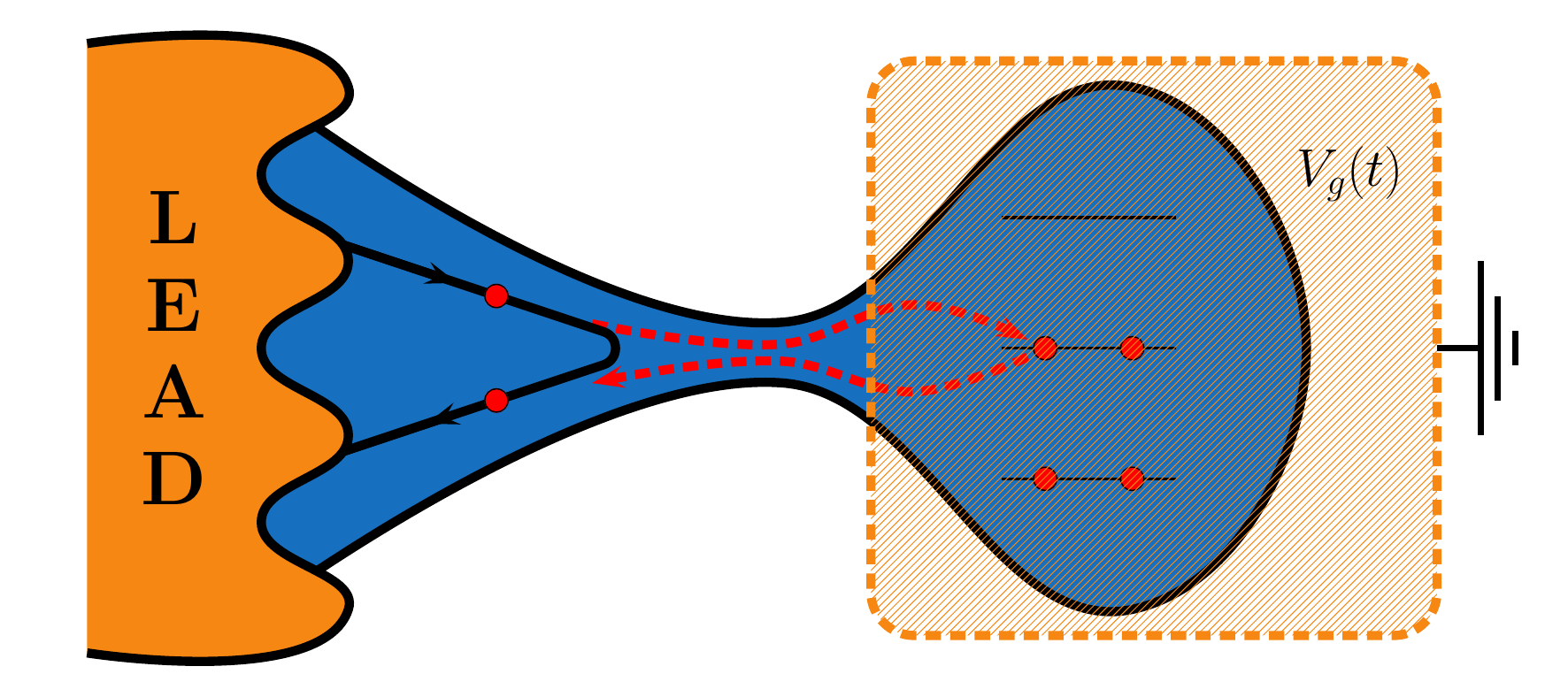}
    \caption{Schematic view of the quantum RC circuit. Electrons coming from a metallic lead can tunnel inside a quantum dot where there are interactions between electrons. An oscillating voltage $V_g(t)$ is applied by a metallic gate coupled capacitively to the dot.}\label{fig:rc}
  \end{figure}

A fundamental and paradigmic experiment~\cite{gabelli2006}
 in the topic of high frequency
transport is the quantum capacitor, or quantum RC circuit.
It consists of a quantum dot attached to a reservoir lead via a
quantum point contact. In the experiment of 
Refs.~\cite{gabelli2006,feve2007,mahe2010}, 
a single spin-polarized
channel of the lead is connected to the quantum dot. In addition,
the quantum dot forms a mesoscopic capacitor with a top metallic gate
as illustrated in Fig.~\ref{fig:rc}.
These resistive and capacitive elements constitute the quantum analog
of the classical RC circuit in series. 
By applying a time-dependent
voltage on the top gate, the quantum capacitor can be operated in
the linear~\cite{gabelli2006} 
or non-linear~\cite{feve2007,mahe2010,parmentier2011}
 regimes, and charge can be transferred between
the dot and the lead alternatively. In the linear regime,
an AC drive changes infinitesimally and periodically 
the charge of the dot. Matching the low frequency admittance of
the dot with the corresponding formula for a classical RC circuit
\begin{equation}\label{admi}
\frac{I(\omega)}{V_g (\omega)} = -i \omega C_0 (
1+ i \omega C_0 \, R_q ) + {\rm O} (\omega^3)
\end{equation}
allows us to define a quantum capacitance $C_0$ and 
a charge relaxation resistance $R_q$. $V_g$ denotes the gate voltage,
$I$ the current from the dot to the lead. 

Recent experiments have
developed an alternative way of measuring the admittance of
a single or double quantum dot by embedding it in a microwave
cavity~\cite{delbecq2011,frey2011}, see also experiments using radiofrequency 
reflectometry~\cite{persson2010,ciccarelli2011,chorley2012}.
A good quality factor greatly enhances the coupling
to the photons at the resonant frequencies of the cavity and
the admittance can be read off the phase of a 
microwave probe. Only measurements of the capacitance~\cite{cottet2011} 
were realized so far but the technique is in principle able to 
capture the charge relaxation resistance.


The term {\it charge relaxation resistance} was coined in the 
seminal work of B\"uttiker, Pr\^etre and 
Thomas~\cite{buttiker1993,buttiker1993b,pretre1996}
 where a general theory
of time-dependent coherent transport was put forward. Coulomb
interactions were treated using a discrete RPA-like model,
in which the screening of the potential imposed by the gate
is taken into account self-consistently. One particular outcome
of this theory is the prediction of a quantized and universal
resistance $R_q=h/2 e^2$ in the case of a polarized single-channel
lead. This remarkable universality was notably confirmed
 experimentally~\cite{gabelli2006}. 
For more than one channel,  $R_q$ was expressed
in terms of statistical distribution of dwell times, corresponding
in that case to Wigner-Smith delay times~\cite{gopar1996,brouwer1997}, 
and computed for chaotic or weakly disordered quantum dots using 
random matrix theory 
techniques~\cite{brouwer1997b,pedersen1998,buttiker1999,buttiker2005}.
Interestingly, the RPA-like screening approximation emerges as the
leading order contribution in an $1/N$ expansion ($N$ being
the number of channels in the lead connected to the dot).
This expansion was devised~\cite{Brouwer2005}
 as a general method to describe the
interplay of coherent transport and interaction in quantum dots
driven out of equilibrium. Other aspects of the linear AC response
of the quantum RC circuit were theoretically addressed including 
the charge relaxation resistance and the 
inductive response~\cite{wang2007}
within a Luttinger model for a long tube connected 
between electrodes~\cite{blanter1998,pham2003}, the quantum to 
classical transition in the presence of finite temperature
or dephasing probes~\cite{nigg2008,buttiker2009} and the effect
of possibly strong interaction within an 
Hartree-Fock approach~\cite{nigg2006,buttiker2007,Ringel2008}
or by developing a real-time diagrammatic expansion in the
tunnel coupling~\cite{Splettstoesser2010}.
It has also been suggested that the quantum RC circuit could be used
to detect efficiently the state of a nearby double-dot 
system~\cite{nigg2009} or to probe charge fractionalization 
in a quantum spin-Hall insulator~\cite{garate2011}.

The charge relaxation resistance has also been investigated for
small metallic islands where the tunnel junction to the reservoir
is described by a large number of weakly 
transmitting channels~\cite{rodionov2009,Petitjean2009}.
In this regime, a mapping to the problem of a single particle 
on a ring subject to dissipation has been exploited~\cite{etzioni2011}
 to demonstrate
a new fixed point at large transparency associated to the quantized
resistance $R_q = h/e^2$.

Before pursuing our discussion on the linear AC response, let us
mention that the experiment~\cite{gabelli2006} on the quantum RC circuit
 was also driven into a 
non-linear regime~\cite{feve2007,mahe2010,parmentier2011}.
A square-shaped excitation with an amplitude comparable to the level
spacing turns the mesoscopic capacitor into a single electron source.
An electron (and a hole) is thus sent into the lead at each period
of the driving signal. This experiment has initiated an intense
theoretical activity on dynamics and quenches in coherent and interacting
nanoscaled systems~\cite{Moskalets2008,keeling2008,albert2010,albert2011,andergassen2011,kashuba2011,debora2011}.

The prediction of the universal resistance $R_q = h/2 e^2$ for a single
channel was recently reconsidered~\cite{mora2010,hamamoto2010} 
by treating Coulomb interaction in an 
exact manner. In this way, the strong Coulomb blockade regime could be 
addressed analytically, yet at the price of treating the coupling to the lead
perturbatively, for either small or large transparency.
All analytical calculations: bosonization of the 
fermionic degrees of freedom, 
perturbation in the dot-lead coupling and mapping to the 
Kondo model~\cite{glazman1990,matveev1991},
point to the fact that the resistance $R_q = h/2 e^2$ survives arbitrarily 
strong interactions for all transmissions, thereby reinforcing
its universality.
In addition to this result, it was shown~\cite{mora2010}
 that a large dot, with an effectively
vanishing level spacing, also supports a universal, albeit different, charge
relaxation resistance $R_q = h/e^2$. Interactions in the lead, for the edge
state of a fractional quantum Hall state for instance, 
are added~\cite{mora2010,hamamoto2010} for free
in bosonization. They simply renormalize the charge
relaxation resistance $R_q = h/2 \nu e^2$ where $\nu$ denotes the electron
filling factor in the bulk. 
For $\nu < 1/2$, a quantum phase transition occurs~\cite{hamamoto2010} 
as a function of the
transmission (dot-lead coupling) into an incoherent regime where resistance
quantization is lost (see also Ref.~\cite{Furusaki2002} 
where a similar transition was obtained). 
Complementary to these analytical findings, Monte-Carlo calculations~\cite{hamamoto2010} 
have confirmed the universality of the charge relaxation resistance
for $\nu > 1/2$ (including the case of non-interacting leads $\nu=1$) ,
{\it i.e.} for all interactions and transmissions, and the quantum phase
transition for $\nu < 1/2$.

Surprises came from the single-channel case with spinfull 
electrons (in contrast to the fully polarized edge states). 
When a single level on the dot participates to electronic 
transport, the quantum RC circuit is described by
the Anderson model~\cite{hewson1997}. 
At zero magnetic field and small excitation 
frequency, the Korringa-Shiba relation~\cite{shiba1975} 
on the dynamical charge 
susceptibility implies a quantized universal resistance 
$R_q = h/4 e^2$, again in agreement with the original RPA
approach~\cite{nigg2006}. 
Note that this corresponds to a weak charge response and a weak 
low frequency dissipation $\propto C_0^2 R_q$ when the charge on 
the dot is quenched by strong Coulomb interaction 
(for example, in the Kondo regime). 
Finite magnetic fields or higher frequencies do not alter the
freezing of charge fluctuations,
but they allow processes which redistribute the spin 
populations on the dot and cause an increase of energy dissipation. 
The result is a giant peak that the charge relaxation resistance develops
with either frequency or magnetic field~\cite{lee2011}. 
Note that although the peak 
in magnetic field emerges at the Kondo energy scale and therefore 
originates from strong correlations, it does not contradict the 
Fermi liquid nature of the model at low frequency
but arbitrary magnetic field. A Fermi liquid description~\cite{filippone2011}
 is thus able
to reproduce analytically the properties of the peak, showing 
that the peak disappears at the particle-hole symmetric point. 
A generalized Korringa-Shiba relation can be derived that expresses the 
resistance $R_q$ in terms of static susceptibilities.
As we shall discuss in this paper, the Fermi-liquid approach introduced 
in Refs.~\cite{garst2005,filippone2011} 
is in fact quite general and should apply to a variety of models.

To summarize, analytical and numerical calculations~\cite{mora2010,hamamoto2010}
 have proven that the universal resistance $R_q = h/2 e^2$ remains valid 
even for strong
Coulomb interaction on the dot. Nevertheless, the physical reason
for this universality is somehow hidden in the formalism, 
especially in the bosonization approach.
The intent of this paper is to bridge the gap between
the weakly interacting 
model of Refs.~\cite{buttiker1993,buttiker1993b,pretre1996,Ringel2008}
and the strongly interacting approach of Refs.~\cite{mora2010,hamamoto2010},
by proposing a general Fermi liquid framework that captures all interaction 
regimes within a single model, and recovers  $R_q = h/2 e^2$ in the
single-channel case.

The paper is organized essentially along two directions. The first part,
corresponding to Sec.~\ref{sec:fermi}, presents the main ideas underlying the
Fermi liquid approach. Due to the lack of phase-space available 
for inelastic scattering
(this is the standard Landau argument for Fermi 
liquids~\cite{nozieres1974,aleiner1998,luttinger1961}), elastic processes
dominate the physics of interacting quantum dots at low energy~\cite{clerk2001}.
Hence, low energies are described by non-interacting electrons backscattered by
the dot and, using the Friedel sum rule~\cite{*[{   }] [{,  
the precise formulation of the Friedel sum
rule involves the displaced charge and not the charge on the dot,
although the two quantities coincide in the wide-band limit}] langreth1966},
one arrives at the Korringa-Shiba
formula for the dynamical charge susceptibility. While the capacitance
is proportional to the local charge susceptibility,  $C_0=e^2\chi_c$,
the charge relaxation resistance is expressed as a combination of static charge 
susceptibilities given by Eq.~\eqref{resistance}, with $R_q=h/2 e^2$ in the single-channel
case. The discussion is extended to the case of a large dot in 
Sec.\ref{sec:large}, where the dot itself is described as a Fermi liquid of
non-interacting electrons separated from the lead, and the charge relaxation resistance
is in units of $R_q=h/e^2$. Both resistances, $h/2 e^2$ and
$h/e^2$, are direct consequences of the Fermi liquid structure Eq.~\eqref{lowener0}
and of the Friedel sum rule, and are therefore almost independent of the gate
voltage.

\begin{figure}[b]
  \includegraphics[width=8cm]{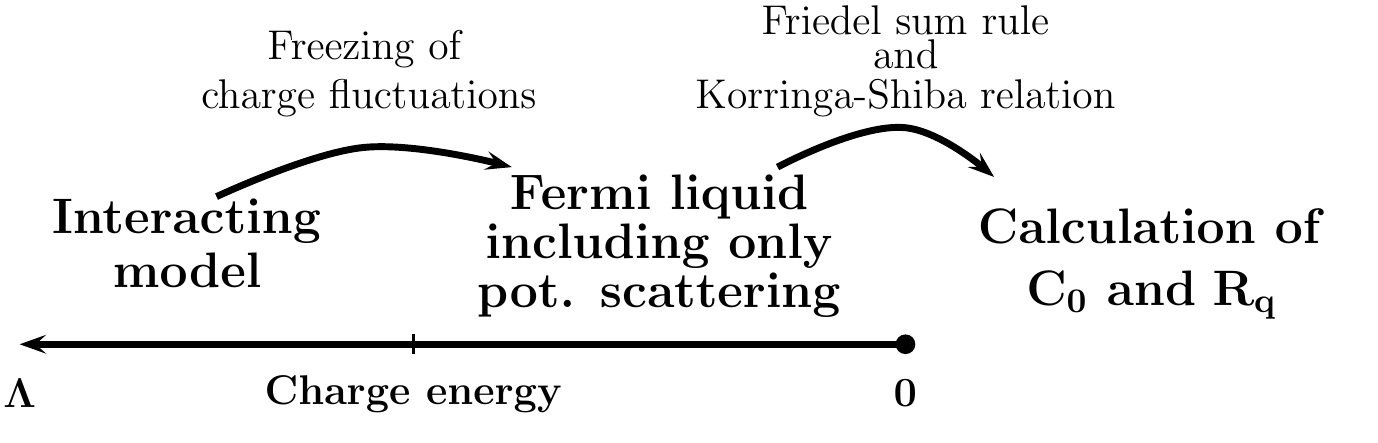}
  \caption{Line of reasoning developed in this paper. The linear response of interacting systems is described by a non-interacting Fermi liquid at low
energy. It is then possible to compute the quantum capacitance $C_0$ and the charge relaxation resistance $R_q$. $\Lambda$ is the cutoff energy scale in the RG approach.}\label{fig:scheme}
\end{figure}

The second part of the paper is detailed in Sec.~\ref{sec:renormalization}.
The validity of the Fermi liquid approach is explored using a 
renormalization-group (RG) 
analysis for two specific models relevant to describe the quantum RC circuit: 
the Anderson model and the Coulomb blockade model.
This is illustrated in Fig.~\ref{fig:scheme}. The perturbative RG
approach allows us, for both models, 
to calculate explicitly the low energy effective
Hamiltonian in agreement with  Sec.~\ref{sec:fermi}.
In particular, the Friedel sum rule is checked by comparing our predictions with
existing results from the literature: Bethe-ansatz calculations for the Anderson model
and a perturbative calculation for the Coulomb blockade model.
In addition for the Anderson model, we provide a rigorous mapping to the Kondo 
model and derive an analytic formula for the charge susceptibility in powers of 
the hybridization out of the particle-hole symmetric point.

\section{Fermi liquid approach}\label{sec:fermi}

\subsection{Hamiltonians}\label{sec:hamiltonians}

The Fermi liquid approach, to be discussed below in the core
of this section, does not rely on a specific Hamiltonian but rather
defines a universality class for the low energy behaviour
of coupled dot and lead systems.
Despite of this, we will introduce two particular Hamiltonians: the
Anderson model~\cite{hewson1997} 
and a second model describing an interacting quantum dot 
with internal levels that we
shall call the Coulomb blockade model (CBM)~\cite{glazman1990,matveev1991}. 
For these two models indeed, we shall see
in Sec.~\ref{sec:renormalization} that they fall into the
category of Fermi liquids at low energy.
Nevertheless, one has to keep in mind that the conclusions
of this section
 are in no way restricted to the models 
Eqs.~\eqref{am},~\eqref{mat} but would also 
be applicable for more complicated interactions on the dot
mixing, for example, long range and short range components.

The quantum RC circuit is described~\cite{lee2011,filippone2011} 
by the Anderson model
when the level spacing in the dot is sufficiently large and electron 
transport is not spin-polarized.
The Hamiltonian takes the form
\begin{equation}\label{am}
\begin{split}
H_{\rm AM} &= \sum_{\sigma,k} \varepsilon_{k\sigma} 
 c^\dagger_{k\sigma} c_{k\sigma}
+  \varepsilon_{d} 
\, \hat{n} \\
& + U \hat{n}_{\uparrow} \hat{n}_{\downarrow}
+ t \sum_{k,\sigma} \left( c_{k\sigma}^\dagger  d_\sigma +  d_\sigma^\dagger
c_{k\sigma} \right),
\end{split}
\end{equation}
with the electron operators $c_{k\sigma}$ and $d_\sigma$ of spin
 $\sigma$ for the lead and the dot respectively.
The lead electrons are characterized by the single-particle 
energies $\varepsilon_k$
with the constant density of states  $\nu_0$. The electron number 
on the dot is $\hat{n} = \hat{n}_{\uparrow} + \hat{n}_{\downarrow}$
with $\hat{n}_\sigma =  d_\sigma^\dagger d_\sigma$.
$U$ is the interaction energy, or the charging energy in the case of
a quantum dot, related to the gate capacitance $C_g$ through $U = e^2/C_g$.
 $\varepsilon_d = - e V_g$ is the single-level energy of the dot.
It is tuned via the electrostatic coupling of the quantum dot to the 
metallic gate. $t$ is the amplitude for electron tunneling between the dot
and the lead. We will later need the hybridization constant 
$\Gamma = \pi \nu_0 t^2$.

The second model, which we call the Coulomb blockade model 
(CBM)~\cite{glazman1990,matveev1991},
 is appropriate for a larger dot with
at least a few energy levels in the dot relevant for transport.
Its simplest version includes spinless electrons and a single
channel in the lead but it can be straightforwardly extended
to $N$ channels. The Hamiltonian splits as~\cite{grabert1994b,mora2010}
$H_{\rm CBM}  = H_0  + H_c + H_T$ with
\begin{subequations}\label{mat}
\begin{align}
\label{h0} H_0 = \sum_{k,\sigma} \varepsilon_k & \, 
c^\dagger_{k\sigma} c_{k\sigma} 
+ \sum_{l,\sigma} \varepsilon_l \, d^\dagger_{l\sigma} d_{l\sigma}
+  \varepsilon_{d}  \, \hat{n}, \\[1mm]
\label{inter} H_c  = E_c  \hat{n}^2, & \qquad
 H_T = t \sum_{k,l,\sigma} \left( d^\dagger_{l\sigma} c_{k\sigma} 
+ c^\dagger_{k\sigma} d_{l\sigma} \right),
\end{align}
\end{subequations}
where the three terms describe respectively non-interacting
electrons with single-particle energies $\varepsilon_k$ (lead)
and $\varepsilon_l$ (dot), the charging energy due to strong
Coulomb repulsion in the dot and the tunneling of electrons
between the dot and the lead. $E_c = e^2/2 C_g$ and $\sigma=1,\ldots,N$.
Again $\hat{n} = \sum_\sigma
\hat{n}_\sigma = \sum_{k,\sigma} d^\dagger_{k\sigma} d_{k\sigma} $ 
is the total number of electrons in the dot
and  $\varepsilon_d = - e V_g$ is set by the gate voltage.
The level spacing in the dot is finite but can be sent to zero
for a large enough quantum dot. In this paper, the density of
states $\nu_0$ is chosen for simplicity 
to be the same in the lead and in the dot but none of our results
are affected by releasing this constraint. 
It will be convenient later to use the dimensionless
conductance $g = N \, (\nu_0 \, t)^2$.
For both models, the total number of electrons,
that is a constant of motion, is written $\hat{N}_t$.
 
Although the derivations of Sec.~\ref{sec:renormalization} are valid
only away from charge degeneracy, the Anderson and the 
one-channel Coulomb blockade models are both Fermi liquid
for all gate voltages $\varepsilon_d$. Our effective Fermi liquid 
approach, which predicts the resitances $h/2 e^2$ and $h/e^2$,
 is therefore applicable for all values of $\varepsilon_d$
including the Coulomb peaks at the charge degeneracy points.
In the multi-channel Coulomb blockade model, the Fermi liquid
approach breaks down only right at 
charge degeneracy (non-Fermi liquid fixed points).

\subsection{Non-interacting electrons}

Before addressing the general Fermi liquid approach, that is the central 
part of this paper, it is instructive to shortly review the 
non-interacting case~\cite{buttiker1993,buttiker1993b,pretre1996} 
following the discussion of Ref.~\cite{Ringel2008}.
First the coupling to the gate voltage $\varepsilon_d (t) \hat{n}$ 
can be gauged out by a simple unitary transformation 
$U(t) = e^{i \int^t d t' \varepsilon_d (t') \hat{N}_t}$
shifting all single-particle energies by $- \varepsilon_d$, notably
$\varepsilon_k \to \varepsilon_k-\varepsilon_d$ for the single-particle 
energies in the lead. 

Coherent electrons are described by delocalized 
wavefunctions propagating throughout the quantum dot. Different 
trajectories, corresponding to single or multiple reflections at the 
quantum point contact opening the dot, interfere by 
adding their amplitudes. In the absence of Coulomb interaction, 
electrons behave very much like photons traversing a dispersive medium: 
an energy-dependent phase shift $\Phi(\varepsilon_k-\varepsilon_d)$ is 
accumulated after passing through the dot.
Although the discussion here is meant to be general, we can illustrate 
with a specific example. The phase shift reads
\begin{equation}
e^{i \Phi(\varepsilon)} = \frac{r-e^{i 2 \pi 
\varepsilon/\Delta}}{1-r e^{i 2 \pi \varepsilon/\Delta}}
\end{equation}
for a quantum dot embedded in a quantum Hall edge state, $r$ being the 
reflection coefficient of the quantum point contact, $\Delta$ the level 
spacing in the dot and $2 \pi \varepsilon/\Delta$ the phase accumulated 
after a single turn around the dot.
With this picture in mind, the current from the dot to the lead 
can be computed from the Landauer-B\"uttiker 
scattering formalism~\cite{buttiker1993,buttiker1993b,pretre1996,Ringel2008}. 
At low frequency,
\begin{equation}\label{current}
I (t) =  \frac{e^2}{h} \left[ V_g (t) - V_g (t-\tau) \right],
\end{equation}
$\tau$ is the Wigner-Smith delay time, or the typical dwell time 
of an electron in the dot ($\varepsilon_F$ is the Fermi energy),
\begin{equation}\label{tau}
\tau = \hbar \left.
\frac{d \Phi (\varepsilon-\varepsilon_d)}{d \varepsilon} 
\right|_{\varepsilon = \varepsilon_F}
=  - \hbar \frac{d \Phi (\varepsilon_F-\varepsilon_d)}{d \varepsilon_d} ,
\end{equation}
and Eq.~\eqref{current} holds as long as $\omega \tau \ll 1$.

Eqs.~\eqref{current} and~\eqref{tau} show that a 
non-zero current is an effect of the dispersive cavity. 
Physically, 
a time varying gate voltage $\varepsilon_d (t)$ implies that different 
times of electron arrivals correspond to different energies 
$\varepsilon-\varepsilon_d(t)$ and therefore different
phase shifts. All electrons are not slowed down in the same way and charge
can accumulate either in the dot or in the lead. Fourier 
transform of Eq.~\eqref{current} is compared to Eq.~\eqref{admi} and yields
\begin{subequations}\label{coherent}
\begin{align}
\label{universal1}
R_q & = \frac{h}{2 e^2}, \\[2mm]
\label{capa}
C_0 & = \frac{e^2}{h} \tau,
\end{align}
\end{subequations}
such that $\tau = 2 R_q C_0$. The coherent regime thus gives a quantized
and universal resistance $R_q = h/2 e^2$.

The same reasoning applies to the case of $N$ channels in the lead
connected to the dot. The result for the charge relaxation resistance
reads~\cite{buttiker1999,nigg2006}
\begin{equation}\label{nonintresistance}
R_q = \frac{h}{2 e^2} \, \frac{\sum_{\sigma=1}^N \tau_\sigma^2}
{\left( \sum_{\sigma=1}^N \, \tau_\sigma \right)^2} =
\frac{h}{2 e^2} \, \frac{\sum_{\sigma=1}^N \nu_{0\sigma}^2}
{\left( \sum_{\sigma=1}^N \, \nu_{0\sigma} \right)^2},
\end{equation}
where $\tau_\sigma$ is the Wigner-Smith delay time in the channel
$\sigma$. $\nu_{0\sigma}= \tau_\sigma/h$~\cite{buttiker1999} 
denotes the density of states for channel $\sigma$ in the dot.

\subsection{Physical picture}\label{sec:phys}

Simple arguments can be given to argue
that the results Eq.~\eqref{coherent}
extend to the general case of interacting electrons. Despite its apparent 
simplicity, the expression Eq.~\eqref{tau} 
for the time delay $\tau$ that controls the
low frequency transport, conveys a fundamental message:
only electrons in the immediate vicinity of the Fermi surface participate
to AC transport at low frequency. These electrons however, following the 
conventional Fermi liquid 
argument~\cite{luttinger1961,nozieres1974,aleiner1998}, 
are somehow protected against interactions 
due to the restriction of  available phase-space close
to the Fermi surface. Hence, even in the presence of strong Coulomb 
repulsion on the dot, low energy electrons in the lead behave 
essentially as if they were non-interacting and the line of reasoning 
detailed above generalizes to the interacting case. This 
generalization is possible only in the generic
situation of Fermi liquid behaviour at low energy while non-Fermi liquid
fixed points usually require a delicate tuning of 
coupling constants~\cite{oreg2003,pustilnik2004,potok2007}.
We note in passing that a similar restoration of phase-coherence caused
by Coulomb interaction was explored in a quantum 
dot T-junction geometry~\cite{clerk2001}.

Relatedly, Eq.~\eqref{capa} for the capacitance or the static response 
of the dot, coincides with an exact expression as we shall now 
argue. In the interacting case, the phase shift $\delta (\varepsilon_d) 
= \Phi(\varepsilon_k-\varepsilon_d)/2$
(there is factor $2$ difference between the phase shift definition used 
in the non-interacting case for instance in 
Refs.~\cite{gabelli2006,Ringel2008}, and the conventional
phase shift of the Friedel sum rule~\cite{langreth1966}) is related to 
the occupancy of
the dot via the Friedel sum rule $\delta / \pi = \langle \hat{n} \rangle$.
Substituted in Eq.~\eqref{capa}, one finds 
\begin{equation}\label{capa2}
C_0 = - 2 \pi \frac{\partial \langle \hat{n} \rangle}{\partial \varepsilon_d}
\, \frac{e^2}{2 \pi} = e^2 \chi_c,
\end{equation}
where $\chi_c$ is the static charge susceptibility of the dot. This result
coincides with the mere definition of $C_0$ if we recall that 
$I = \partial_t \, e \langle \hat{n} \rangle$ and consider the
time-integrated version of Eq.~\eqref{admi} in the static limit.

Let us now clarify an important point: 
the similarities between the scattering properties
of non-interacting and interacting electrons do not imply that interactions
have no effect as they can strongly renormalize the energy dependence
of the phase shift. For non-interacting electrons, a comparison
between Eq.~\eqref{capa} and Eq.~\eqref{capa2} suggests that
$\tau /h = \nu_{d0} = \chi_c$, clearly a sensible result.
  Indeed, for free electrons, 
the density of states represents the number 
of one-particle states that fall below the Fermi
surface upon a unit increase of the Fermi energy or, 
equivalently, a unit decrease of the reference energy 
$\varepsilon_d$. The Pauli principle then
implies that it is also the number of added electrons,
that is the charge susceptibility.
However, in the presence of interactions between electrons,
$ \nu_{d0} \ne \chi_c$ in general. $\chi_c$ is sensitive only
to charge excitations while all excitations
contribute to the density of states $\nu_{d0}$.
This difference is exemplified~\cite{lee2011,filippone2011}
 by the Anderson model 
in the Kondo regime where the local density of states  
exhibits a peak at the Fermi energy, mostly due to
spin-flip excitations, whereas the charge
susceptibility is suppressed by Coulomb blockade~\cite{kawakami1990}.

\subsection{Effective model and Korringa-Shiba formulas}\label{sec:effec}

We begin by some general remarks on dissipation for the
quantum RC circuit.
Integrating Eq.~\eqref{admi} with respect to time, we find the 
low frequency expansion~\cite{mora2010,filippone2011}
\begin{equation}\label{expansion}
\frac{e^2 \langle \hat{n} (\omega) \rangle}
{- \varepsilon_d (\omega)} = C_0 + i \omega C_0^2 R_q
+ {\cal O} (\omega^2),
\end{equation}
for the charge on the dot in the presence of a time-dependent gate
voltage. The comparison with standard linear response theory allows
for deriving the correspondences
\begin{equation}\label{corres}
\chi_c (\omega=0) = \frac{C_0}{e^2}, \qquad
\left. {\rm Im} \chi_c (\omega) \right|_{\omega \to 0}
= \frac{ \omega C_0^2 R_q}{e^2},
\end{equation}
with the dynamical charge susceptibility
\begin{equation}
\chi_c (t-t') =  \frac i\hbar  \theta (t-t') \langle
[ \hat{n} (t), \hat{n} (t') ] \rangle.
\end{equation}
Following standard linear response theory, the
 power dissipated in the presence of an AC drive of the gate voltage,
$\varepsilon_d (t) = \varepsilon_d^0 + \varepsilon_\omega \cos \omega t$ with 
$\varepsilon_\omega$ small enough to be in the linear regime,
is given quite generally, and specifically in the 
models Eqs.~\eqref{am} and~\eqref{mat}, by
\begin{equation} \label{dissip}
{\cal P} = \frac{1}{2} \varepsilon_\omega^2 \, \omega 
\, {\rm Im} \chi_c (\omega).
\end{equation}
The only requirement for Eq.~\eqref{dissip} to hold is that
$\varepsilon_d$ appears in the Hamiltonian only through
the term $\varepsilon_d \, \hat{n}$.

Having exposed in Sec.~\ref{sec:phys} 
the physical reasons for which the universal resistance is
insensitive to arbitrary interaction in the dot, we proceed 
and develop a low energy effective model that shall 
prove Eq.~\eqref{universal1} explicitly.  
General arguments are sufficient to reconstruct the structure of 
the low energy effective Hamiltonian. Nevertheless one has to keep 
in mind that the whole discussion that follows is based on the 
assumption of a Fermi liquid infrared (IR) fixed point.

The first piece of the Hamiltonian is a free part $H_0 = \sum_k
\varepsilon_k c_k^\dagger c_k$. 
It describes the bulk lead electrons. The dot 
having a finite spatial extension, it is not able to alter the 
bulk properties (like the Fermi velocity or the effective mass) 
of the lead electrons.
 Operators perturbing the free term $H_0$ can be 
classified, in the RG sense, according to their 
relevance. Again the finite size of the dot implies that these 
operators involve only the field operators $\psi ({\bf r}=0) 
= \sum_k c_k$, ${\bf r}=0$ being the entrance 
of the dot or the position of the quantum point contact, $\psi^\dagger (0)$
and derivatives thereof. There is only a single marginal operator, 
all other operators being irrelevant. Keeping the former and 
discarding the latter, the low energy Hamiltonian assumes the form
\begin{equation}\label{lowener0}
H = \sum_{k} \varepsilon_k c^\dagger_{k} c_{k}
+ K (\varepsilon_d) \sum_{k,k'} c^\dagger_{k} c_{k'}.
\end{equation}
The second term in this Hamiltonian describes a structureless 
scattering potential  
($\propto \delta ({\bf r})$) placed at the dot-lead boundary 
or entrance of the dot. For the two initial models 
Eqs.~\eqref{am} and~\eqref{mat},
the Friedel sum rule relates the phase shift of this scattering 
potential to the mean occupation of the dot~\cite{langreth1966}, namely
\begin{equation}\label{occupation}
\langle \hat{n} \rangle = - \frac{1}{\pi} \arctan \left[ \pi \nu_0
K (\varepsilon_d) \right].
\end{equation}
Drawing on the ideas of Refs.~\cite{garst2005,filippone2011}, 
the strategy that we shall adopt to compute the charge 
relaxation resistance is the following: the power dissipated 
by the gate voltage 
$\varepsilon_d (t) = \varepsilon_d^0 + \varepsilon_\omega \cos \omega t$ 
in the linear regime, given by Eq.~\eqref{dissip}, can also be computed
from the low energy model Eq.~\eqref{lowener0}. The identification
between the general expression Eq.~\eqref{dissip} and the low energy
calculation allows us to derive
a Korringa-Shiba formula for ${\rm Im} \chi_c (\omega)$ which
further determines $R_q$.

Let us now start from the low energy Hamiltonian and expand 
it to first order with respect to $\varepsilon_\omega$ ($K_0 =
 K(\varepsilon_d^0)$),  
\begin{equation}\label{lowener1}
\begin{split}
H = & \sum_{k} \varepsilon_k  c^\dagger_{k} c_{k}
+ K_0 \sum_{k,k'} c^\dagger_{k} c_{k'} \\[2mm]
& + K'(\varepsilon_d^0) \, \varepsilon_\omega \cos \omega t
\, \sum_{k,k'} c^\dagger_{k} c_{k'}.
\end{split}
\end{equation}
The scattering by the potential $K_0 \delta ({\bf r})$ is a single-particle
problem which can be readily diagonalized. Rewriting the Hamiltonian in
terms of the corresponding scattering states, characterized 
by the fermionic operators $\tilde{c}_k$,
absorbs the second term  into the first one in Eq.~\eqref{lowener1}.
 Additionally, this change of basis introduces a multiplicative 
constant~\cite{krishna1980a,*krishna1980b} in the 
third term of Eq.~\eqref{lowener1} and 
\begin{equation}\label{lowener2}
H = \sum_{k} \varepsilon_k  \tilde{c}^\dagger_{k}  \tilde{c}_{k}
+  \frac{K'(\varepsilon_d^0) \, \varepsilon_\omega \cos \omega t}
{1+(\pi \nu_0 K_0)^2}  \, \sum_{k,k'} \tilde{c}^\dagger_{k} \tilde{c}_{k'}.
\end{equation}
This last expression is conveniently written in terms of the static
susceptibility 
$\chi_c = - \partial \langle \hat{n} \rangle/\partial \varepsilon_d$,
which is obtained through the derivative of Eq.~\eqref{occupation}
\begin{equation}\label{occupation2}
\chi_c = \frac{\nu_0 \, K'(\varepsilon_d^0)}{1+(\pi \nu_0 K_0)^2} ,
\end{equation}
such that Eq.~\eqref{lowener2} becomes
\begin{equation}\label{lowener3}
H = \sum_{k} \varepsilon_k  \tilde{c}^\dagger_{k}  \tilde{c}_{k}
+  \frac{\chi_c}{\nu_0} \varepsilon_\omega 
\cos \omega t \, \, \sum_{k,k'} \tilde{c}^\dagger_{k} \tilde{c}_{k'}.
\end{equation}
It can be checked for consistency that this low energy model 
satisfies the Friedel sum rule in the static limit $\omega \to 0$. 
The scattering potential in Eq.~\eqref{lowener3} adds the phase shift 
$\delta_1 = - \pi \nu_0  \frac{\chi_c}{\nu_0} \varepsilon_{\omega}$ to 
lead electron wavefunctions. 
The Friedel sum rule then translates this phase shift
into a shift in the occupation number on the dot 
\begin{equation}
\delta \langle \hat{n} \rangle = \delta_1 /\pi 
= - \chi_c \, \varepsilon_\omega,
\end{equation}
in agreement with the above definition
of the static charge susceptibility.

Now that we have derived a more compact low energy Hamiltonian, 
we rely again on linear
response theory in order to compute the power dissipated 
upon exciting the gate voltage sinusoidally. 
It involves the operators $\hat{A} = (\chi_{c} /\nu_0) 
\sum_{k,k'} \tilde{c}^\dagger_{k} \tilde{c}_{k'}$
coupled to the AC drive in Eq.~\eqref{lowener3} or
\begin{equation} \label{dissip2}
{\cal P} = \frac{1}{2} \varepsilon_\omega^2 \, \omega 
\, {\rm Im}  \chi_{\hat{A}} (\omega),
\end{equation}
where  $\chi_{\hat{A}\sigma} (t-t') = \frac{i}{\hbar} \theta (t-t') \langle
[ \hat{A}_\sigma (t), \hat{A}_\sigma (t') ] \rangle$.
Perhaps unsurprisingly, energy dissipation occurs 
at low energy through the production
of single electron-hole excitation. ${\rm Im} \chi_A (\omega)$ 
is easily computed at zero temperature for non-interacting 
electrons (first term in Eq.~\eqref{lowener3}), 
${\rm Im}  \chi_{\hat{A}} (\omega) = \hbar \pi \chi_{c}^2 \omega$,
similar to the result of a Fermi golden rule calculation.
Comparing Eq.~\eqref{dissip} and Eq.~\eqref{dissip2}, 
we obtain the Korringa-Shiba formula~\cite{shiba1975}
\begin{equation}\label{korringa}
{\rm Im} \chi_c (\omega) = 
\hbar \pi \omega \, \chi_{c}^2.
\end{equation}
This result, substituted in Eq.~\eqref{corres}, recovers the universal charge 
relaxation resistance $R_q = h/2 e^2$. 

So far, our analysis has concentrated on the single channel
case but its generalization to $N$ channels is straightforward. 
One then finds a generalized Korringa-Shiba formula
\begin{equation}\label{korringa2}
{\rm Im} \chi_c (\omega) = 
\hbar\pi \omega \, \sum_{\sigma=1}^N \chi_{c\sigma}^2.
\end{equation}
The static susceptibilities 
$\chi_{c\sigma} =  - \partial \langle \hat{n}_\sigma
\rangle / \partial \varepsilon_d$
measure the sensitivity of the occupations 
of the dot, for each channel $\sigma$, to a change in the gate 
voltage. The charge relaxation 
resistance thus takes the form
\begin{equation}\label{resistance}
R_q = \frac{h}{2 e^2} \, \frac{\sum_{\sigma=1}^N \chi_{c\sigma}^2}
{\left( \sum_{\sigma=1}^N \, \chi_{c\sigma} \right)^2}
\end{equation}
and resembles very much the non-interacting one Eq.~\eqref{nonintresistance}.
They coincide after the identification $\nu_{0\sigma}=\chi_{c\sigma}$, already
discussed in Sec.\ref{sec:phys} and valid only for free electrons.
Therefore  Eq.~\eqref{resistance}
 gives the correct generalization of Eq.~\eqref{nonintresistance}
 to the interacting case.

\subsection{Effective model for a large dot}\label{sec:large}

The case of a large quantum dot deserves a specific discussion. 
By a large dot, we mean that the single-particle spectrum can be 
treated as continuous in the dot
such that energy dissipation takes place both in the lead and in the dot.

In fact, finite dots have a finite level spacing $\Delta$ and one may wonder
at which energy scale the spectrum can be considered as continuous.
One solution, proposed in Ref.~\cite{mora2010}, is to send an AC signal
with a bandwidth larger than $\Delta$ in order to smear the
discreteness of the spectrum.
In that case, the frequency $\omega$ has to be larger than $\Delta$.
A second possibility is to use a frequency $\omega$ larger than
the energy $\sqrt{\Delta E_{\rm TH}}$, where $E_{\rm TH} > \Delta$
is the Thouless energy, or inverse time of diffusion through the dot.
Above this energy, it has been shown~\cite{altshuler1997} 
that the one-particle density of states
loses its discreteness due to electron-electron interactions.

The Fermi liquid picture still applies to the large dot with the subtlety
that dot and lead constitute two separate Fermi liquids.  
At low energy,
the transfer of electrons between the dot  and the lead is energetically
prohibited~\cite{aleiner1998} and electrons are effectively fully backscattered 
at the boundary between the dot and the lead. The effective model 
for the lead is the same as above (Eq.~\eqref{lowener0})
with the potential scattering strength related to the mean occupation
of the dot via Eq.~\eqref{occupation}.
 
At first glance, it may seem that the dot simply adds
an additional channel for dissipation so that Eq.~\eqref{resistance} 
would apply with $N=2$.
However, as we shall see below, there exists a lead/dot symmetry 
which reestablishes universality 
in the charge relaxation resistance. 
We first note that the fact that the charging energy is 
ascribed to the dot is physically sensible but mathematically 
arbitrary. Using the fact that the total number 
of electrons in the system $\hat{N}_t$ is conserved by 
the Hamiltonian, one can replace $\hat{n} = \hat{N}_t 
- \hat{n}_L$, $\hat{n}_L$ being the number of electrons
in the lead, and transfer the Coulomb interaction to the lead. 
Therefore, the low energy model for the dot is the same as 
for the lead, namely Eq.~\eqref{lowener0}, but the strength
of the scattering potential for dot electrons, 
noted $K_{\rm dot} (\varepsilon_d)$, is now given by
\begin{equation}\label{occupation3}
\hat{N}_t 
- \langle \hat{n}
\rangle = \langle \hat{n}_L \rangle = 
- \frac{1}{\pi} \arctan \left[ \pi \nu_0
K_{\rm dot} (\varepsilon_d) \right].
\end{equation}
An alternative formulation of the same physics is that the phase shift 
accumulated after backscattering at the boundary is 
$\delta (\varepsilon_d)$ for lead electrons and 
$\delta_t - \delta (\varepsilon_d)$ for dot electrons, 
where $\delta_t = \hat{N}_t/\pi$.
Following the same steps as in Sec.~\ref{sec:effec}, one finds the effective 
low energy Hamiltonian
\begin{equation}\label{lowener4}
\begin{split}
H = & \sum_{k,\alpha=L/D}  \varepsilon_k  \tilde{c}^\dagger_{k\alpha}  
\tilde{c}_{k\alpha} \\[2mm]
&+  \frac{\chi_c}{\nu_0} \varepsilon_\omega 
\cos \omega t \, \, \sum_{k,k'} \left( \tilde{c}^\dagger_{kL} \tilde{c}_{k'L}
- \tilde{c}^\dagger_{kD} \tilde{c}_{k'D} \right),
\end{split}
\end{equation}
where $L/D$ stands for lead/dot electrons. The Korringa-Shiba formula is then
\begin{equation}\label{korringa3}
{\rm Im} \chi_c (\omega) =  2 \hbar\pi \omega \, \chi_{c}^2
\end{equation}
and the charge relaxation resistance $R_q = h/e^2$.
The extension to $N$ channels is straightforward.

\section{Renormalization}\label{sec:renormalization}

\subsection{Outline}

The aim of this section is to justify in more detail the low
energy form Eq.~\eqref{lowener0} for the Anderson Eq.~\eqref{am}
and CBM Eq.~\eqref{mat} models.
We recall that the discussion of Sec.~\ref{sec:effec} rests
on two fundamental assumptions: (i) an infrared (IR)  Fermi
liquid fixed point and (ii) the Friedel sum rule Eq.~\eqref{occupation}
applies. The two models will be discussed separately.

\begin{figure}[h!]
  \includegraphics[width=8.5cm]{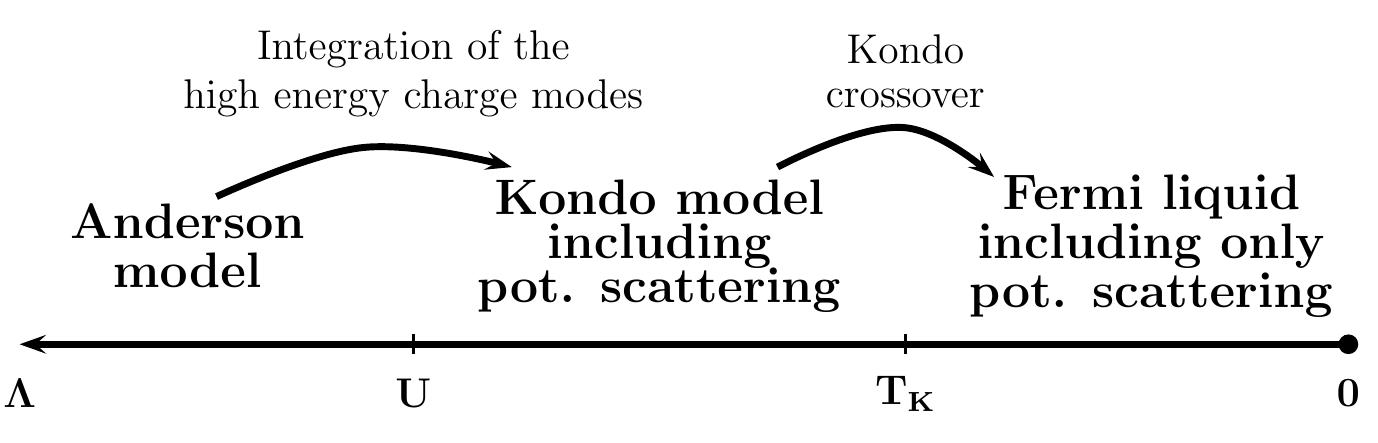}
  \caption{Renormalization-group (RG) analysis of the Anderson model.}\label{fig:schemaam}
\end{figure}

The Friedel sum rule and the Fermi liquid properties at low energy
are well-established for the Anderson 
model~\cite{langreth1966,haldane1978}. Our motivation is thus
a practical one: we calculate the scattering potential
$K(\varepsilon_d)$ perturbatively to second order in $\Gamma$
in the Kondo regime. To this end, a RG treatment
is first carried out on the Hamiltonian Eq.~\eqref{am}
and stopped at intermediate energies $T_K \ll \Lambda \ll U$,
where $\Lambda$ denotes the running energy scale (or cutoff).
The RG approach remains perturbative up to these
energies and can be performed explicitly.
 At this stage, charge excitations
have been completely integrated out and the Kondo model is obtained,
with exchange and scattering potential terms.
Proceeding towards lower energies, the RG procedure
becomes non-perturbative across
the Kondo temperature $T_K$ and we thus rely on the work of Cragg
and Llyod~\cite{cragg1978,cragg1979a,cragg1979b}. 
They showed that, whereas the exchange term
disappears at low energy, leading simply to a $\pi/2$ scattering
phase shift,
the scattering potential is unaffected by the Kondo crossover up to
small corrections that are negligible within our second order
calculation. The main steps of the RG procedure up to the IR
fixed point are summarized 
in Fig.~\ref{fig:schemaam}.  
We finally obtain the scattering potential
\begin{equation}\label{finalK}
\begin{split}
\nu_0 K(\varepsilon_d) &= -\frac{\Gamma}{2\pi}\frac{2\varepsilon_d+U}
{\varepsilon_d(\varepsilon_d+U)}\left[1-\frac{\Gamma \, U}
{\pi \varepsilon_d(\varepsilon_d+U)} \right. \\[1mm]
& \left. -\frac{\Gamma \, U^2}
{\pi \varepsilon_d(\varepsilon_d+U)(2\varepsilon_d+U)}
\ln\left( \frac{\varepsilon_d+U}{-\varepsilon_d} \right) \right].
\end{split}
\end{equation}
This expression can be used to compute the static charge
susceptibility $\chi_c$ at zero temperature,
\begin{widetext}
\begin{equation}\label{chied}
\begin{split}
\chi_c=\frac{\Gamma}\pi\left\{\frac1{(\varepsilon_d+U)^2}+\frac1{\varepsilon_d^2}+\frac{2\Gamma}\pi\right.&\left.\left[\frac1{(\varepsilon_d+U)^3}-\frac1{
\varepsilon_d^3 } \right ]
+\right.\\
+&\left.\frac\Gamma\pi\left[\left(\frac1{\varepsilon_d+U}-\frac1{\varepsilon_d}\right)^3+2\left(\frac1{\varepsilon_d+U}-\frac1{\varepsilon_d}\right)\left(\frac1{
\varepsilon_d^2}-\frac1{(\varepsilon_d+U)^2}\right)\ln\frac{\varepsilon_d+U}{-\varepsilon_d}\right]\right\},
\end{split}
\end{equation}
\end{widetext}
in agreement with a  Bethe-ansatz calculation at the particle-hole
symmetric point~\cite{horvatic1985}.
Let us finally mention that the preliminary perturbative
renormalization (stopped at $\Lambda \gg T_K$)
recovers Haldane's formula~\cite{haldane1978,haldane1978a} 
for the Kondo temperature of the Anderson model, namely 
\begin{equation}\label{tkhaldane}
T_K = \frac{e^{1/4+C}}{2 \pi} \, 
\sqrt{\frac{2 \Gamma U}{\pi}} \, e^{{\pi \varepsilon_d(\varepsilon_d +U)/2 U \Gamma}},
\end{equation}
where $C=0.5772$ 
is the Euler constant. $e^{1/4+C}/(2 \pi) = 0.364$ is in agreement
with Ref.~\cite{haldane1978a}.

The situation is different for the  Coulomb blockade model which 
is known to be a Fermi liquid at low energy, 
except close to the charge degeneracy for $N \ne 1$~\cite{lebanon2003}.
For this model, perturbative renormalization can be
performed explicitly down to low energy. In this paper, this 
is done to second order in the conductance $g$, 
{\it i.e.} to
fourth order in the tunneling matrix element $t$. We
are thus able to check the scenario described in Sec.~\ref{sec:large}
of two separated Fermi liquids at low energy with scattering
phase shifts in agreement with the Friedel sum rule.
To leading order in a large $N$ calculation, we find 
for the backscattering of lead electrons at the boundary
\begin{equation}\label{finalKmat}
K(\varepsilon_d ) = \nu_0t^2\ln\frac{E_c+\varepsilon_d}{E_c-\varepsilon_d}+N\nu_0^3t^4\left(A[\varepsilon_d]-A[-\varepsilon_d]\right),
\end{equation}
with $A[\varepsilon_d]$ reported in \eqref{A}. Translated
into the dot occupancy with the help of the Friedel sum rule,
this result coincides with a direct calculation 
by Grabert \cite{grabert1994a,grabert1994b}.

Below we detail the perturbative calculations of the scattering 
potential $K(\varepsilon_d)$ leading to Eq.~\eqref{finalK} 
and Eq.~\eqref{finalKmat}. A refined quantum field theory approach
is not necessary for the leading order (first order in $\Gamma$) and
we simply use the unitary Schrieffer-Wolff transformation
to compute $K(\varepsilon_d)$. This is done in Sec.~\ref{sec:poor}.
In Sec.~\ref{sec:anderson} and~\ref{sec:matveev}, 
the next order is obtained within a more general field theoretical 
approach to renormalization.

\newcommand{\comm}[2]{\left[#1,#2\right]}
\newcommand{\ket}[1]{\left|#1\right>}
\newcommand{\bra}[1]{\left<#1\right|}
\newcommand{\av}[1]{\left<#1\right>}

\subsection{Schrieffer-Wolff unitary transformation}\label{sec:poor}

When the charging energy, $U$ or $E_c$, largely exceeds the hybridization
to the lead due to tunneling, the different charge states
on the dot become well-separated in energy. For temperatures
$T \ll U,E_c$, one charge state defines the low energy sector,
the others being only virtually occupied.
The Schrieffer-Wolff unitary transformation~\cite{schrieffer1966}
accounts for these virtual states by integrating them perturbatively
into an effective Hamiltonian acting in the low energy sector.
The Schrieffer-Wolff transformation was initially devised for the 
Anderson model, it shall be applied in this paper 
also to the Coulomb blockade model Eq.~\eqref{mat}.

\subsubsection{Anderson model}\label{sec:ander}

We focus on the Kondo regime defined by $0<-\varepsilon_d < U$,
that is in-between the Coulomb peaks, with
$|\varepsilon_d|/\Gamma \gg \ln(U/\Gamma)$ in order to neglect
the renormalization~\cite{haldane1978} in the position of these
peaks.
The low energy sector then corresponds to a single electron on the dot.
The tunneling term in the Anderson model~\eqref{am}, that we call $H_T$, 
couples subsequent charge states and is linear in $t$.
The idea of the Schrieffer-Wolff unitary transformation $e^{i S}$
is to cancel this linear tunnel coupling, 
thereby producing couplings between
charge states that have higher orders in $t$. This strategy
is realized with the choice
\begin{equation}\label{swprescription}
i H_T = [S, H_{\rm AM}^0],
\end{equation}
where $H_{\rm AM}^0$ is Eq (\ref{am}) with $t=0$. The rotated Hamiltonian assumes the form
\begin{equation}\label{rotham}
H_{\rm AM}' = e^{iS} H_{\rm AM} e^{-iS} = 
H_{\rm AM}^0 + \frac{i}{2} \, [S, H_T].
\end{equation}
After this transformation and
specifically for the Anderson model in the Kondo regime, 
the coupling of the single-charge
sector to other charge states starts as $t^3$ at most. 
Hence, for the purpose of a calculation up to second order in $t$,
the single-charge sector decouples from 
other charge states. The resulting effective Hamiltonian
is the Kondo model which includes a potential scattering term
\begin{equation}\label{kondo}
 H_{\rm AM}' =H_0
+J \mathbf S\cdot\mathbf s+K \sum_{kk'\sigma}c^\dagger_{k\sigma}c_{k'\sigma},
\end{equation}
where $\mathbf S$ denotes the spin operator for 
the single electron in the dot and $\mathbf
s=\sum_{kk'\sigma\sigma'}c^\dagger_{k\sigma}\frac{{\bm 
\tau}_{\sigma\sigma'}}2c_{k'\sigma'}$ the local spin of lead electrons.
${\bm \tau}_{\sigma\sigma'}$ is the vector composed of the Pauli matrices. 
The coupling constants are given by
\begin{subequations}\label{couplingconstants}
\begin{align}
\nu_0 J = \nu_0 J_0=& \frac{2 \Gamma}{\pi}
\left(\frac1{\varepsilon_d+U}-\frac1{\varepsilon_d}\right),\label{jpm}\\
\nu_0 K = \nu_0 K_0=&-\frac{\Gamma}{2 \pi}
\left(\frac1{\varepsilon_d+U}+\frac1{\varepsilon_d}\right).\label{wpm}
\end{align}
\end{subequations}
Notice that $K=0$ at the particle-hole symmetric point $\varepsilon_d=-U/2$, 
leaving exclusively the Kondo interaction with $J=8t^2/U$.
In deriving the Kondo model Eq.~\eqref{kondo}, one discards the
energy dependence of the coupling constants 
Eqs.~\eqref{couplingconstants}  because the model Eq.~\eqref{kondo}
 is applicable only for energies much smaller than the charging energy
$\varepsilon \ll U,|\varepsilon_d|$.
The Schrieffer-Wolff transformation is thus an economical way of
renormalizing the Anderson model up to the energy scale
$\Lambda$ with $T_K \ll \Lambda \ll U,|\varepsilon_d|$.

Controlling the low energy behavior, 
the IR fixed point ($\Lambda \to 0$) of the Kondo model with 
potential scattering
Eq.~\eqref{kondo}, has been identified by
Cragg and Llyod using a combination of analytical 
and numerical calculations~\cite{cragg1978,cragg1979a,cragg1979b}.
It is a Fermi liquid in which lead electrons at the Fermi level
acquire a phase shift with three contributions: $\pi/2$, the phase shift 
corresponding to $K(\varepsilon_d)$ and a correction proportional
to $\Gamma^3$ that we can legitimately neglect in our first order calculation
and even in the second order calculation of Sec.~\ref{sec:anderson}.
Absorbing the $\pi/2$ phase shift into a redefinition of the lead electrons,
we indeed obtain the low energy form Eq.~\eqref{lowener0} where 
$K(\varepsilon_d)$ is given by Eq.~\eqref{wpm}.

With the help of the Friedel sum rule Eq.~\eqref{occupation}, 
the dot occupancy is computed from Eq.~\eqref{wpm}
and then the static charge susceptibility
at the particle-hole symmetric point reads 
\begin{equation}
  \chi_c=-\left.\frac{\partial\av{\hat n}}{\partial\varepsilon_d}\right|_{\varepsilon_d=-U/2}
=\frac{8\Gamma}{\pi U^2},
\end{equation}
in agreement with a Bethe ansatz calculation expanded to leading order 
in $\Gamma/U$~\cite{horvatic1985}.

\subsubsection{Coulomb blockade model}\label{submat}

In the Coulomb blockade model of Eq.~\eqref{mat},
the tunnel coupling between the subsequent charge states is also linear in $t$
and the principle of the  Schrieffer-Wolff transformation 
remains the same as in Sec.~\ref{sec:ander}
with the choice Eq. (\ref{swprescription}) and the rotated Hamiltonian Eq. (\ref{rotham}) where
the subscript AM is replaced by CBM.

Following Grabert~\cite{grabert1994a,grabert1994b}, 
we define the operator $\hat{n}$, which gives
the number of electrons in the dot, as being independent 
from the fermionic degrees of freedom $d_{l\sigma}$ and we note $\ket n$ the charge
state with $n$ electrons. In this representation, the tunneling term in Eq.~\eqref{mat}
reads
\begin{equation}\label{grabert}
  H_T=t\sum_{n,k,l}\left(d^\dagger_lc_k\ket{n+1}\bra n+c^\dagger_kd_l\ket {n-1}\bra n\right).
\end{equation}
For the sake of simplicity, but with no loss of generality, we restrict
ourselves here to the single-channel case $N=1$. The operator $S$ that generates the unitary transformation, solution of Eq. (\ref{swprescription}), 
takes the form $s+s^\dagger$ where
\begin{subequations}
\begin{align}
  s=&it\sum_{k,l,n} s_{kln} c^\dagger_k d_l \ket{n-1} \bra n,\\[2mm]
  s_{kln}=&\frac1{\varepsilon_l-\varepsilon_k+E_C(2n-1)+\varepsilon_d}.
\end{align}
\end{subequations} 
In contrast with the Anderson model, a $t^2$ coupling between each
state $\ket n$ with $\ket{n\pm2}$ is produced by the Schrieffer-Wolff transformation.
Nonetheless, it is possible to remove this coupling
 by applying a second unitary transformation
whose specific form will not be needed here. After that, the Hamiltonian becomes block
diagonal in the charge states up to second order in $t$.  
 For $-E_c < \varepsilon_d < E_c$, the charge state $n=0$ defines
the low energy sector and the rotated Hamiltonian is $H'_{\rm
CBM} = H_0 + H_{\rm B}$,
\begin{equation}\label{swgrabert}
H_{\rm B} = \frac{t^2}2\sum_{kk'll'}
  \left(s_{kl0}d^\dagger_{l'}c_{k'}c^\dagger_kd_l-s_{kl1}c^\dagger_kd_ld^\dagger_{l'}c_{k'} 
+ {\rm h.c.}\right).
\end{equation}
The normal ordering of operators is necessary to classify them according
to their RG relevance. For example,
\[
d^\dagger_l c_kc^\dagger_{k'} d_{l'}=\delta_{ll'}\theta(-\varepsilon_l)c_kc^\dagger_{k'}+\delta_{kk'}\theta(\varepsilon_k)d^\dagger_l d_{l'} + : d^\dagger_l c_kc^\dagger_{k'} d_{l'} :
\]
where $\theta (\varepsilon)$ is the Heaviside function and $:\ldots:$ denotes
normal ordering with respect to the Fermi sea. The last term in this expression
describes interaction between lead and dot electrons; it is irrelevant 
at low energy and can be discarded. The two quadratic terms give marginal terms
in the Hamiltonian corresponding to backscattering of electrons at the lead-dot
boundary. If we take for instance the first one, its contribution to $H_B$ is
\begin{equation}
\begin{split}
& \qquad \frac{t^2}2\sum_{kk'} c_{k'} c^\dagger_k \sum_{ll'} \delta_{ll'}\theta(-\varepsilon_l)
s_{kl0}  \\[1mm]
&= \sum_{kk'} c_{k'} c^\dagger_k \frac{\nu_0 t^2}2 \int_{-D_0}^0
\frac{d \varepsilon_l}{\varepsilon_l-\varepsilon_k-E_c+\varepsilon_d} \\[1mm]
& = \sum_{kk'}  c^\dagger_k c_{k'} \frac{\nu_0 t^2}2 
\ln \left( \frac{\varepsilon_k+E_c-\varepsilon_d+D_0}{\varepsilon_k+E_c-\varepsilon_d}
\right).
\end{split}
\end{equation}
Since we consider the low energy properties of the model, it is consistent 
to take $\varepsilon_k=0$
in this expression, the difference being again irrelevant in the RG sense.
Collecting all marginal terms in Eq.~\eqref{swgrabert} and performing the
integrals, we obtain the effective low energy model
to leading order in $t$
\begin{equation}\label{effintmat}
H_{\rm CBM}' = H_0+  \frac{g}{\nu_0} \ln\left(\frac{E_c-\varepsilon_d}{E_c+\varepsilon_d}\right)\left[\sum_{ll'}d^\dagger_l d_{l'}-\sum_{kk'}c^\dagger_kc_{k'}\right],
\end{equation}
in which the intermediate cutoff $D_0$ has disappeared~\cite{matveev1991}
and the dimensionless conductance $g= (\nu_0 \, t)^2$ has been introduced.
This expression confirms the scenario proposed in Sec.~\ref{sec:large}
for a large dot in which the two Fermi liquids, lead and dot, completely 
separate at low energy while they experience potential scattering terms 
with opposite amplitudes.

The phase-shift acquired by lead electrons from Eq.~\eqref{effintmat} is
\begin{equation}
  \delta=\pi g \, \ln\left(\frac{E_c-\varepsilon_d}{E_c+\varepsilon_d}\right)
\end{equation}
and, applying the Friedel sum rule, one finds the dot mean occupancy
\begin{equation}\label{swmat}
 \av {\hat n}=\frac\delta\pi= g \, \ln
\left(\frac{E_c-\varepsilon_d}{E_c+\varepsilon_d}\right),
\end{equation}
in agreement with previous direct perturbative 
computations~\cite{matveev1991,grabert1994a,grabert1994b} to order $g$.
The above result Eq.~\eqref{swmat} is unchanged for $N$ channels as long
as $g= N (\nu_0 \, t)^2$.
~\\

As we shall show below, the different conclusions obtained in this section
carry over to the next order in  $g$.
We were not able to iterate the Schrieffer-Wolff transformation to next orders.
We shall instead use a field theoretical approach to derive the renormalization
of the two models to second order in $\Gamma$ and $g$, 
the Anderson model in  Sec.~\ref{sec:anderson}
and the CBM in Sec.~\ref{sec:matveev}.

\subsection{Field theory approach I\\ Renormalization of the Anderson model}\label{sec:anderson}

Quantum Field Theory is an efficient tool for carrying out 
RG calculations and extracting the low-energy behavior.
Standard diagrammatic perturbation techniques are not 
applicable to the Anderson model
when expanded in the tunneling constant $t$, because of the non-quadratic
interaction term $U \hat{n}_{\uparrow} \hat{n}_{\downarrow}$.
\begin{figure}[htb]
  \includegraphics[width=8cm]{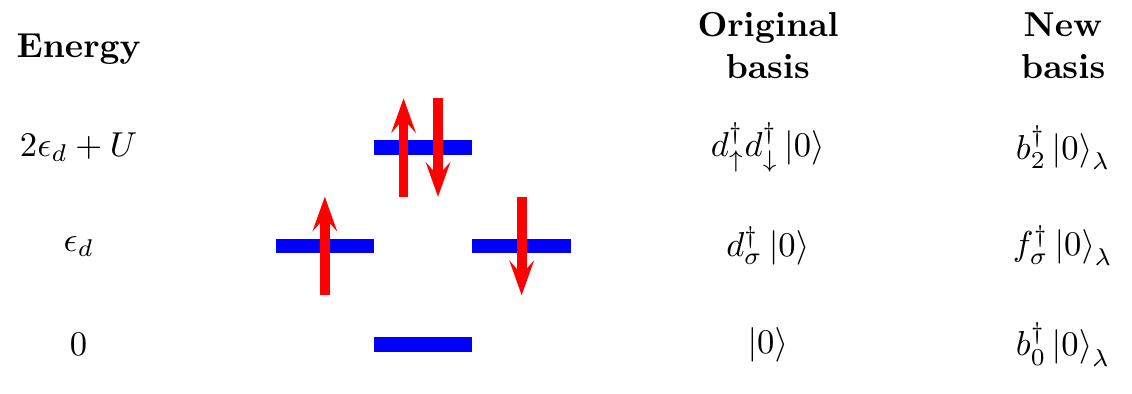}
  \caption{New definition of the states on the dot in Barnes' representation. The energies are 
given for an isolated dot.}\label{fig:barnes}
\end{figure}
To circumvent this problem, we follow a representation due to Barnes~\cite{barnes1976} and
introduce two slave-boson operators $b_0$ and $b_2$, corresponding to the states of the dot
with zero and two electrons respectively. This also necessitates to define
two parafermions operators $f_\sigma$ destroying electrons
on the dot. The relation between the old and the new basis of states
on the dot is summarized in Fig. \ref{fig:barnes} where $\ket{0}_\lambda$
denotes the vacuum state of slave-bosons and parafermions.
The use of slave-bosons in this paper differs substantially from most works
based on slave-bosons. In fact, we are interested in the weak coupling
regime (above the Kondo temperature) and not in describing the Kondo
crossover.
The precise identification between operators is given by
\begin{equation}
d_\sigma = b^\dagger_0 \, f_\sigma+\sigma  f^\dagger_{-\sigma} \, b_2,
\end{equation}
ensuring the anticommutation relation $d^\dagger_\uparrow d^\dagger_\downarrow
+d^\dagger_\downarrow d^\dagger_\uparrow=0$.

Clearly this operation enlarges the Hilbert space by adding unphysical states.
The set of physical states is thus recovered by imposing the constraint
\begin{equation}\label{constraint}
b_2^\dagger b_2 + b_0^\dagger b_0 + \sum_\sigma \, f_\sigma^\dagger f_\sigma = 1,
\end{equation}
which commutes with the Hamiltonian. In this basis, the 
Hamiltonian takes the form $H_{\rm AM} = H_0 + H_c + H_T$, where
\begin{subequations}\label{andpara}
  \begin{align}
  H_0&=\sum_{k\sigma} \varepsilon_kc^\dagger_{k\sigma}c_{k\sigma}+(\varepsilon_d+\lambda)\sum_\sigma f^\dagger_\sigma f_\sigma,\\[1mm]
  H_C&=\lambda b^\dagger_0 b_0+(2\varepsilon_d+U+\lambda)b^\dagger_2 b_2,\\[2mm]
  \begin{split}
  H_T&=t\sum_{k\sigma}\left(c^\dagger_{k\sigma}b^\dagger_0 f_\sigma+f^\dagger_\sigma b_0  c_{k\sigma}\right)\\
  &\quad +t\sum_{k\sigma}\sigma\left(c^\dagger_{k\sigma} f^\dagger_{-\sigma}b_2+b^\dagger_2 f_{-\sigma}c_{k\sigma}\right).
  \end{split}
 \end{align}
\end{subequations}
The chemical potential  $\lambda$ has been introduced in order to eliminate
non-physical states by imposing the constraint Eq.~\eqref{constraint}.
This projection is realized by setting $\lambda \to +\infty$ at the end of
calculations~\cite{abrikosov1970,barnes1976,rodionov2009}.
The key point of the representation of Eq.~\eqref{andpara} is that the
Hamiltonian is quadratic in the absence of tunneling and standard 
diagrammatic techniques become applicable.
To make more easily contact with the Kondo model, it is convenient to
shift the chemical potential $\lambda \to \lambda -\varepsilon_d$.

With the free propagators
\begin{subequations}
\begin{align}
\label{green1}  G^{-1}_{k\sigma}(i\omega_n) =&i\omega_n-\varepsilon_k,\\
\label{green2}  F^{-1}_\sigma(i\omega_n)=&i\omega_n-\lambda,\\
  F^{-1}_0(i\nu_n)=&i\nu_n+\varepsilon_d -\lambda,\\
  F^{-1}_2(i\nu_n)=&i\nu_n-\varepsilon_d-U-\lambda,
\end{align}
\end{subequations}
where $i\omega_n=(2\pi n+1)/\beta$ denote always fermionic Matsubara frequencies
and $i\nu_n=2\pi n/\beta$  bosonic ones,
the action corresponding to Eq.~\eqref{andpara} reads
\begin{widetext}
\begin{equation}
\begin{split}
S=&-\sum_{i\omega_nk\sigma}c^\dagger_{k\sigma}(i\omega_n)G^{-1}_{k\sigma}(i\omega_n)c_{k\sigma}(i\omega_n)-\sum_{i\omega_n\sigma}f^\dagger_\sigma(i\omega_n)F^{-1}_\sigma(i\omega_n)
f_\sigma(i\omega_n)\\[1mm]
&-\sum_{i\nu_n}\left(b^\dagger_0(i\nu_n)F^{-1}_0(i\nu_n)b_0(i\nu_n)+b^\dagger_2(i\nu_n)F^{-1}_2(i\nu_n)b_2(i\nu_n)\right)\\[1mm]
&+\frac t{\sqrt\beta}\sum_{\stackrel{i\omega_ni\nu_n}{k\sigma}}\left(c^\dagger_{k\sigma}(i\omega_n)b^\dagger_0(i\nu_n)f_\sigma(i\nu_n+i\omega_n)+\sigma c^\dagger_{k\sigma}(i\omega_n)f^\dagger_{-\sigma}(i\nu_n-i\omega_n)b_2(i\nu_n)+\mbox{c.c.}\right).
\end{split}
\end{equation}
The action being quadratic in the high energy modes $b_0$ and $b_2$, 
their integration is straightforward and gives an action describing a Kondo model with frequency dependent couplings 
\begin{subequations}\label{skondo}
  \begin{align}
    S'=S_0+\frac1\beta\sum_{\stackrel{kk'\sigma\sigma'\tau\tau'}{i\omega_1i\omega_2i\nu_n}}\left[\mathcal J\mathbf S_{\sigma\sigma'}\cdot\mathbf
s_{\tau\tau'}+\mathcal
K\delta_{\sigma\sigma'}\delta_{\tau\tau'}\right]&c^\dagger_{k\sigma}(i\omega_1)c_{k'\sigma'}(i\omega_2)f^\dagger_\tau(i\nu_n+i\omega_2)f_{\tau'}
(i\nu_n+i\omega_1),\\
\mathcal J=-2t^2\left[F_2(i\nu_n+i\omega_1+i\omega_2)+F_0(i\nu_n)\right],&~~~~\mathcal K=\frac {t^2}2\left[F_2(i\nu_n+i\omega_1+i\omega_2)-F_0(i\nu_n)\right],\label{coupcon} 
  \end{align}
  \end{subequations}
\end{widetext}
where $S_0$ stands for the free action of lead electrons and parafermions. 
\begin{figure}[h]
  \includegraphics[width=8cm]{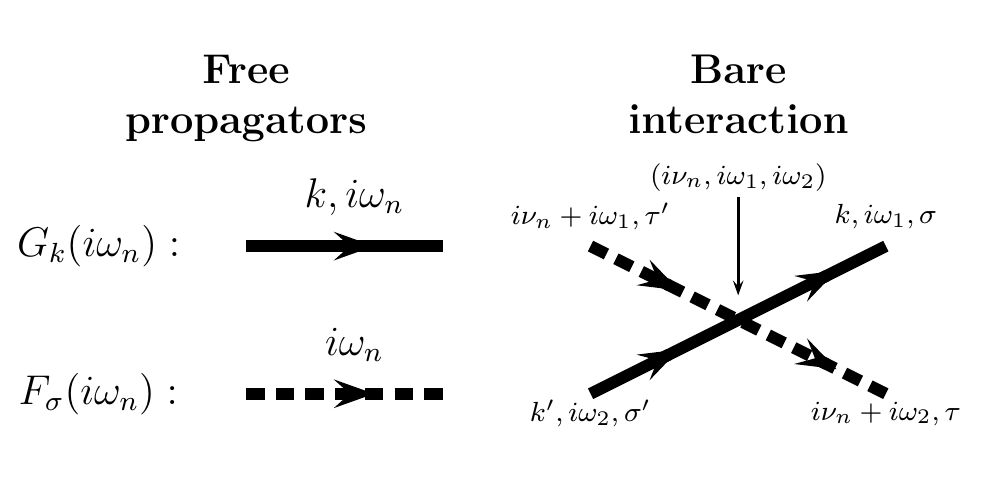}
  \caption{(left) Free 
propagators of the theory . (right) Bare interaction of the theory. 
The frequencies inside the parenthesis are the arguments of
the frequency-dependent couplings  
in (\ref{coupcon}).}\label{fig:interaction}
\end{figure}

\subsubsection{From Anderson to Kondo}\label{anderkondo}

We now elaborate on the connection between the action~\eqref{skondo}, strictly
equivalent to the Anderson model, and the Kondo model.
The infrared (IR) limit of the frequency dependent couplings Eq.~\eqref{coupcon}
is fixed by the poles of the Green's functions, Eq.~\eqref{green1} at the Fermi 
momentum and~\eqref{green2}, {\it i.e.} $i \nu_n = \lambda$, $i\omega_{1} = i \omega_2=0$.
In this limit, the couplings $\mathcal J$ and $\mathcal K$ reproduce exactly
the values $J$ and $K$ of Eqs.~\eqref{couplingconstants} 
obtained after the Schrieffer-Wolff transformation, and they 
control operators that are marginal in the RG sense.
Expanding the couplings in the frequencies $i \nu_n$, $i \omega_{1/2}$ 
around the IR limit generates irrelevant operators in the action
 that we neglect for the moment. The action now takes the form 
\begin{equation}\label{actkondo}
S = S_0 + \int d \tau \left[
J\mathbf S (\tau) \cdot\mathbf s (\tau)+ K\sum_{kk'\sigma}c^\dagger_{k\sigma} (\tau)
c_{k'\sigma} (\tau) \hat{n}_f (\tau) \right],
\end{equation}
with the definitions ${\bf S} = \sum_{\sigma\sigma'}f^\dagger_{\sigma} \frac{{\bm 
\tau}_{\sigma\sigma'}}2f_{\sigma'}$ and $\hat{n}_f = \sum_\sigma f^\dagger_{\sigma}
f_{\sigma'}$ (${\bf s}$ is the local spin operator for lead
electrons defined in Sec.~\ref{sec:ander}).
The limit $\lambda \to +\infty$ now enforces $\hat{n}_f = 1$ and,
using the spin representation due to Abrikosov~\cite{abrikosov1970},
we recognize the action Eq.~\eqref{actkondo} as corresponding
exactly to the Kondo Hamiltonian given by Eq.~\eqref{kondo}.
We thus recover the same result as the Schrieffer-Wolff transformation in Sec.~\ref{sec:ander}.

The irrelevant frequency dependences in the couplings Eq.~\eqref{coupcon} 
 die out upon
lowering the cutoff $\Lambda$ well below $U,\varepsilon_d$
and do not change the low energy Kondo form Eq.~\eqref{actkondo}.
However, along the transient region where they still exist,  they 
can weakly renormalize the value of the constants $J$ and $K$.
This generates a perturbative expansion of $J$ and $K$ in powers
of $\Gamma/U$ where the leading order is given by 
the Schrieffer-Wolff results of Eqs.~\eqref{couplingconstants}.

In order to derive this expansion,  we compute the renormalized
vertex ${\cal V}^R (\Lambda) = {\cal Z} (\Lambda) {\cal V} (\Lambda)$  
where ${\cal V} (\Lambda)$ is the bare vertex and ${\cal Z} (\Lambda)$
the quasiparticle weight of the parafermion propagator. 
Both quantities, to be defined more precisely
below, depend on the running energy scale $\Lambda$.
From works on the multiplicative RG approach to the Kondo model,
it is known that the RG flow in the Kondo model is fully
encoded in the functional dependence of $\mathcal V^R$  
on $\Lambda$ \cite{abrikosov1970,fowler1971,fowler1972}. 
The strategy is thus to compute $\mathcal V^R (\Lambda)$ 
in the Anderson model, or equivalently from the action Eq.~\eqref{skondo},
at intermediate energy $T_K \ll \Lambda \ll U, |\varepsilon_d|$ in 
which case the
functional dependence  of $\mathcal V^R$  
on $\Lambda$ is expected to match the Kondo one.
A comparison of this calculation with the known expression 
for $\mathcal V^R (\Lambda)$
in the Kondo model identifies the values of $J$, $K$ and of the
cutoff $D$ that characterize the effective Kondo Hamiltonian Eq.~\eqref{kondo}
(or Eq.~\eqref{actkondo}) for the Anderson model at intermediate energy.

\subsubsection{The parafermion propagator}

We introduce the parafermion propagator
$\mathcal F_\sigma(\tau-\tau')=-\av{{\cal T}_\tau f_\sigma(\tau)f^\dagger_\sigma(\tau')}$
where ${\cal T}_\tau$ denotes time ordering. In frequency space, the bare 
propagator Eq.~\eqref{green2} is modified by the self-energy
\begin{equation}\label{propagator}
    \mathcal F_\sigma(i\omega_n)= \frac{1}{i\omega_n-\lambda-\Sigma_\sigma(i\omega_n)}.
\end{equation}
Following Solyom's prescription \cite{solyom1974}, the  vertex ${\cal V} (\Lambda)$
is  calculated for equal incoming and outgoing frequencies together with the analytical
continuation
\begin{subequations}\label{ancontinu}
  \begin{align}
    i\omega&\rightarrow 0 & \mbox{for lead electrons,}\\
    i\omega_\Lambda&\rightarrow\tilde\varepsilon_d-\Lambda 
& \mbox{for parafermions.}\label{continuation}
  \end{align}
\end{subequations}
$\Lambda$ is the running energy scale of the RG flow that plays the
role of an IR cutoff in vertex calculations. 
$\tilde \varepsilon_d$ is the renormalized single-level
energy of the dot obtained from the pole of the parafermion 
propagator Eq.~\eqref{propagator}, or 
\begin{equation}\label{singlelevel}
  \tilde \varepsilon_d=\lambda + {\rm Re} \Sigma(\tilde{\varepsilon}_d),
\end{equation}
while the residue at the pole defines the quasiparticle weight ${\cal Z} (\Lambda)$.
 Its dependence on $\Lambda$ can be neglected, ${\cal Z} (\Lambda) \simeq
{\cal Z}_0$, 
as long as $\Lambda \ll {\rm max} (-\varepsilon_d, U+\varepsilon_d)$.
This last assumption is nevertheless only valid within the precision of our 
perturbative calculation (second order in $\Gamma$) where no IR 
divergence appears 
in  $ {\cal Z} (\Lambda)$.
Close to its pole, the parafermion propagator takes the form
\begin{equation}\label{propagator2}
    \mathcal F_\sigma(i\omega_n)= \frac{{\cal Z}_0}{i\omega_n-\tilde \varepsilon_d},
\end{equation}
with ${\cal Z}_0 =[1- \partial_\omega \Sigma(\tilde \varepsilon_d)]^{-1}$.

\begin{figure}
  \includegraphics[width=6cm]{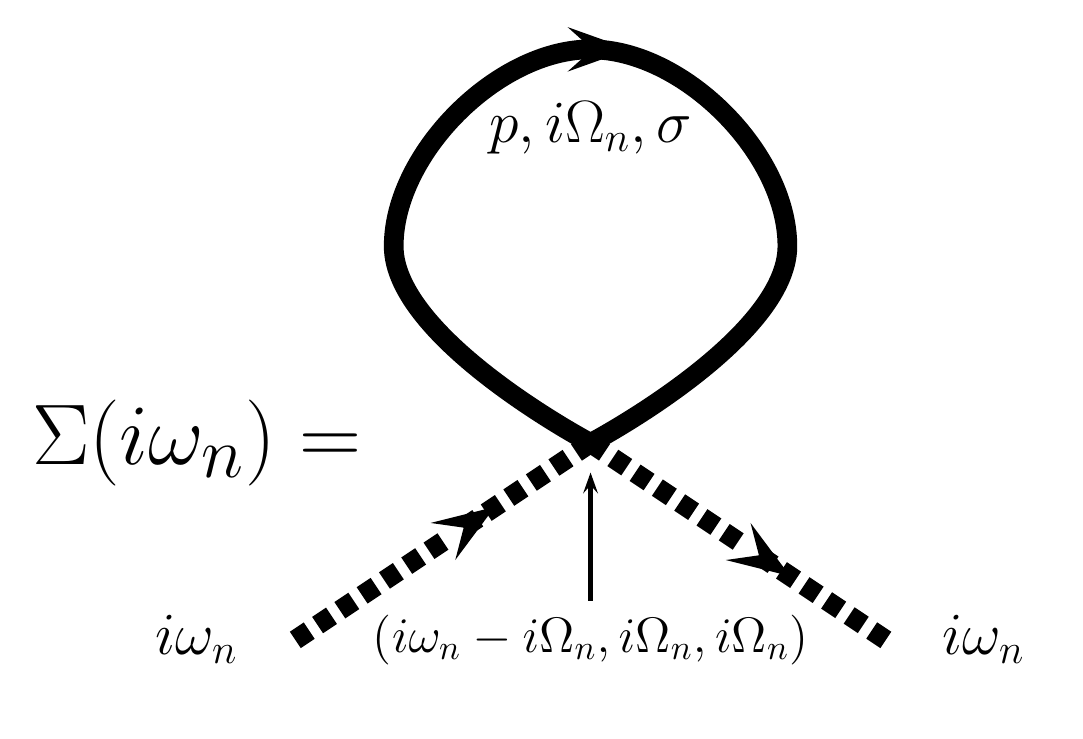}
  \caption{Self-energy of the parafermion propagator to first order $\Gamma$.}\label{fig:self}
\end{figure}

The self-energy $\Sigma_\sigma(i\omega_n)$ is computed to first order in $\Gamma$
to be consistent with our overall second order calculation.
The corresponding diagram is shown in Fig. \ref{fig:self} with the expression
\begin{equation}
\begin{split}
 &  \Sigma(i\omega_n)=\frac{2}{\beta}\sum_{k,i\Omega_n}G_k(i\Omega)\mathcal K(i\omega_n-i\Omega_n,i\Omega_n,i\Omega_n)\\ 
&=t^2\sum_k\left[\frac{f(\varepsilon_k)+b(\varepsilon_d+U+\lambda)}{i\omega_n+\varepsilon_k-\varepsilon_d-U-\lambda}-\frac{f(\varepsilon_k)+b(-\lambda+\varepsilon_d)}{
i\omega_n-\varepsilon_k-\lambda+\varepsilon_d}\right].
\end{split}
\end{equation}
The limit $\lambda\rightarrow\infty$ is taken and the result reads,
after summation over the energies of the lead electrons,
\begin{equation}\label{sigma}
  \Sigma(i\omega_n)=\frac{\Gamma}{\pi}
\left(\ln\frac{\varepsilon_d+U+\lambda-i\omega_n}{D_0}+\ln\frac{\lambda-\varepsilon_d-i\omega_n}{D_0}\right).
\end{equation}
An intermediate cutoff noted $D_0$ has been introduced here 
to remove ultraviolet (UV) divergences. 
However, as we shall see, $D_0$ plays no role and disappears from the final results.
Eq.~\eqref{singlelevel} for $\tilde \varepsilon_d$ can be solved perturbatively.
To first order in $\Gamma$, one finds
\begin{equation}\label{pole}
  \tilde \varepsilon_d=\lambda+\frac{\Gamma}{\pi}
\left(\ln\frac{\varepsilon_d+U}{D_0}+\ln\frac{-\varepsilon_d}{D_0}\right).
\end{equation}
After standard analytical continuation, 
the quasiparticle weight is also extracted 
from Eq.~\eqref{sigma}
\begin{equation}\label{quasiparticle}
  \mathcal Z_0=1-\nu_0t^2\left(\frac1{\varepsilon_d+U}-\frac1{\varepsilon_d}\right)
=1-\frac{\nu_0}2J_0.
\end{equation}
$J_0$ is the Schrieffer-Wolff result Eq.~\eqref{jpm} for $J$.

\subsubsection{The vertex}
\begin{figure}[htb]
  \includegraphics[width=8cm]{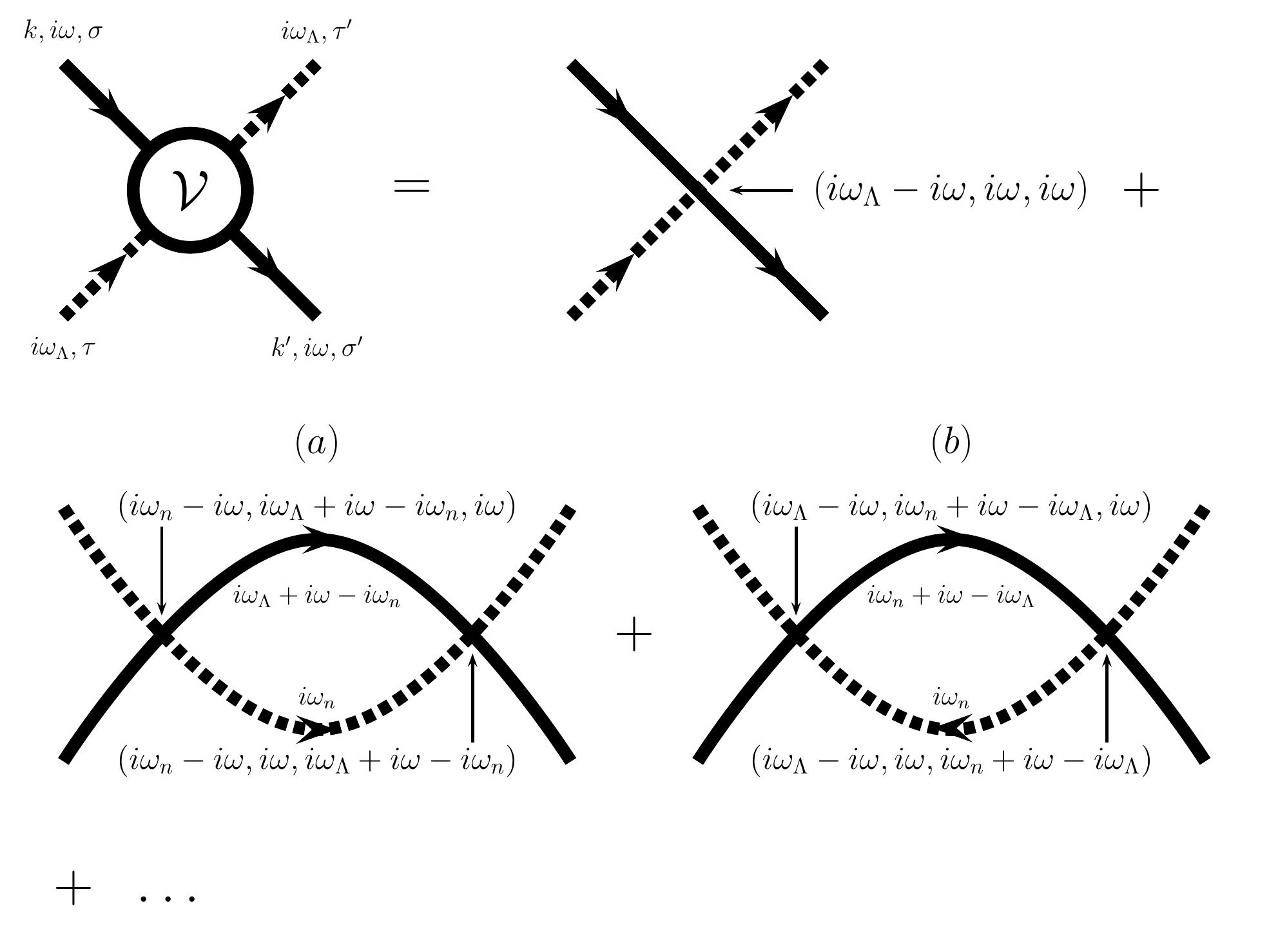}
  \caption{Diagrammatic series for the vertex $\mathcal V$.}\label{fig:vertex}
\end{figure}

The vertex ${\cal V} (\Lambda)$ is given by the series of irreducible diagrams drawn
in Fig. \ref{fig:vertex}.
The calculation is performed up to second order in $\Gamma$, or fourth order
in $t$. The leading order is the bare vertex shown in Fig.~\ref{fig:interaction}
which, after the analytical continuation of Eqs.~\eqref{ancontinu}, reproduces 
the Schrieffer-Wolff expression
\begin{equation}\label{firstorder}
{\cal V} (\Lambda) = 
 \mathcal V_1=\mathbf S_{\tau'\tau}\cdot\mathbf 
s_{\sigma'\sigma}J_0+\delta_{\tau\tau'}\delta_{\sigma\sigma'}K_0,
\end{equation}
with the coupling constants of Eqs.~\eqref{jpm} and~\eqref{wpm}.
In order to derive Eq.~\eqref{firstorder}, we have used that $\tilde \varepsilon_d=\lambda$
to leading order and discarded
$\Lambda/{\rm max} (-\varepsilon_d, U+\varepsilon_d)$ corrections.

The renormalized vertex $\mathcal V^R (\Lambda)$ splits as
\begin{equation}
  \mathcal V^R=\mathcal V_1+\mathcal V_Z+\mathcal V^a+\mathcal V^b,
\end{equation}
where $\mathcal V_Z$ and $\mathcal V^{a,b}$ are second order terms
in $\Gamma$. $\mathcal V_Z$ collects the corrections to $\mathcal V_1$ in
the bare vertex of Fig.~\ref{fig:interaction} brought by
the quasi-particle weight ${\cal Z}_0$ of Eq.~\eqref{quasiparticle} and 
the renormalization of the single-level energy Eq.~\eqref{pole}.
It contains exchange and potential scattering terms
\begin{equation}\label{vz}
\begin{split}
   \mathcal V_Z=&\mathbf S\cdot \mathbf
s\nu_0\left[\left(\frac{J_0^2}4+4K_0^2\right)\left(\ln\frac{-\varepsilon_d(\varepsilon_d+U)}{D_0^2}\right)-\frac{J_0^2}2\right] \\[1mm]
&+\delta_{\tau\tau'}\delta_{\sigma\sigma'}\nu_0\frac{J_0K_0}2\left[\ln\frac{-\varepsilon_d(\varepsilon_d+U)}{D_0^2}-1\right].
\end{split}
\end{equation}
The vertex corrections $\mathcal V^{a,b}$ are illustrated in Fig. \ref{fig:vertex}.
The first one (a) takes the form
\begin{widetext}
\begin{equation}
  \begin{split}
  \mathcal V^a&=-\frac{1}\beta\sum_{k,i\omega_n}F_\sigma(i\omega_n)G_k(i\omega_\Lambda+i\omega-i\omega_n)\left[\mathbf S_{\beta\tau}\cdot\mathbf
s_{\alpha\sigma}\mathcal J_{i\omega_n-i\omega,i\omega_\Lambda+i\omega-i\omega_n,i\omega}+\delta_{\beta\tau}\delta_{\alpha\sigma}\mathcal
K_{i\omega_n-i\omega,i\omega_\Lambda+i\omega-i\omega_n,i\omega}\right]\\
  &\times\left[\mathbf S_{\tau'\beta}\cdot\mathbf s_{\sigma'\alpha}\mathcal
J_{i\omega_n-i\omega,i\omega,i\omega_\Lambda+i\omega-i\omega_n}+\delta_{\beta\tau'}\delta_{\alpha\sigma'}\mathcal
K_{i\omega_n-i\omega,i\omega,i\omega_\Lambda+i\omega-i\omega_n}\right].
\end{split}
\end{equation}
The overall minus comes from diagrammatic rules and the Einstein convention
is used for spin summation. The calculation of this vertex correction is carried
out in Appendix \ref{app:vertex}. We find  $\mathcal V^a=\mathcal V^a_J \,\mathbf
S\cdot\mathbf s+\mathcal V^a_K$ with
\begin{subequations}\label{va}
 \begin{align}
\mathcal V^a_J&=-4\nu_0t^4\left[\frac1{(\varepsilon_d+U)^2}\ln\frac{\Lambda}{D_0}+\frac1{\varepsilon_d(\varepsilon_d+U)}\left(\ln\frac{-\varepsilon_d}{D_0}-\ln\frac{
\Lambda}{D_0}\right)\right],\label{vertak}\\
\mathcal V^a_K&=\nu_0t^4\left[\frac{1}{\varepsilon_d^2}\left(1-\ln\frac{-\varepsilon_d}{D_0}+\ln\frac{\Lambda}{D_0}\right)+\frac1{\varepsilon_d(\varepsilon_d+U)}
\left(\ln\frac {-\varepsilon_d}{D_0}-\ln\frac{\Lambda}{D_0}\right)+\frac1{(\varepsilon_d+U)^2}\ln\frac{\Lambda}{D_0}\right].\label{vertap}
\end{align}
\end{subequations}
The calculation of the (b) diagram follows exactly the same steps
\begin{equation}
\begin{split}
  \mathcal V^b&=-\frac{1}\beta\sum_{k,i\omega_n}F_\sigma(i\omega_n)G_k(i\omega_n+i\omega-i\omega_\Lambda)\left[\mathbf S_{\tau'\beta}\cdot\mathbf
s_{\alpha\sigma}\mathcal J_{i\omega_\Lambda-i\omega,i\omega_n+i\omega-i\omega_\Lambda,i\omega}+\delta_{\beta\tau'}\delta_{\alpha\sigma}\mathcal
K_{i\omega_\Lambda-i\omega,i\omega_n+i\omega-i\omega_\Lambda,i\omega}\right]\\
  &\times\left[\mathbf S_{\beta\tau}\cdot\mathbf s_{\sigma'\alpha}\mathcal J_{i\omega_\Lambda-i\omega,i\omega,i\omega_n+i\omega-i\omega_\Lambda}
+\delta_{\beta\tau}\delta_{\alpha\sigma'}\mathcal K_{i\omega_\Lambda-i\omega,i\omega,i\omega_n+i\omega-i\omega_\Lambda}\right],
\end{split}
\end{equation}
and also contains exchange and potential scattering terms 
\begin{subequations}\label{vb}
\begin{align}
\mathcal V^b_J&=-4\nu_0t^4\left[\frac1{\varepsilon_d^2}\ln\frac{\Lambda}{D_0}+\frac1{\varepsilon_d(\varepsilon_d+U)}\left(\ln\frac{\varepsilon_d+U}{D_0}-\ln\frac{\Lambda}{D_0}\right)\right],\\
  \mathcal V^b_K&=-\nu_0
t^4\left[\frac1{(\varepsilon_d+U)^2}\left(1+\ln\frac{\Lambda}{D_0}-\ln\frac{\varepsilon_d+U}{D_0}\right)+\frac1{\varepsilon_d(\varepsilon_d+U)}\left(\ln\frac{\varepsilon_d+U}
{D_0}-\ln\frac{\Lambda}{D_0}\right)+\frac1{\varepsilon_d^2}\ln\frac{\Lambda}{D_0}\right].
\label{vb2}
\end{align}
\end{subequations}
\end{widetext}
Adding the results from Eqs.~\eqref{firstorder}, \eqref{vz}, \eqref{va} and \eqref{vb}
the renormalized vertex $\mathcal V^R (\Lambda) = \mathcal V_J (\Lambda) \, 
\mathbf S\cdot\mathbf s+\mathcal V_K (\Lambda)$ expands as
\begin{subequations}\label{renormvertex}
\begin{align}
  \mathcal V_J&=J_0 - \frac{\nu_0}2J_0^2-\nu_0J_0^2\ln\left(\frac{\Lambda}
{\sqrt{-\varepsilon_d(\varepsilon_d+U)}}\right),\label{jver}\\
  \mathcal V_K&=K_0 + \frac{\nu_0}2J_0 \, K_0
+\frac{\nu_0}8 J_0^2 \ln \left(
\frac{\varepsilon_d+U}{-\varepsilon_d}\right).\label{wver}
\end{align}
\end{subequations}
As anticipated, $D_0$ has disappeared from these final expressions
and the charging energy $ \sqrt{-\varepsilon_d(\varepsilon_d+U)}$ acts
as an effective high-energy cutoff.
Remarkably, the IR cutoff  $\Lambda$ also disappears from the expression
of the potential scattering term  $\mathcal V_K$ where the limit $\Lambda \to 0$  can safely
be taken. However, the $\ln \Lambda$ dependence in $ \mathcal V_J$ signals
the onset of the Kondo singularity which develops at low energy and
restricts the validity of Eqs.~\eqref{renormvertex} to the 
energy window $T_K \ll \Lambda \ll \sqrt{-\varepsilon_d(\varepsilon_d+U)} $.

In agreement with the leading order calculation of Sec.~\ref{sec:ander},
$ \mathcal V_K$  vanishes at the particle-hole symmetric point  $\varepsilon_d=-U/2$.
 This property is
in fact expected by symmetry to hold to all orders in $\Gamma$.
%
\subsubsection{Kondo temperature and charge susceptibility}

In order to test our predictions Eqs.~\eqref{renormvertex}, we compare
them to existing results from the literature. 
We show that Eqs.~\eqref{renormvertex} give
access for the Anderson model to the Kondo temperature and  the static 
charge susceptibility. Our expression for the Kondo temperature
reproduces a standard result due to Haldane and the charge susceptibility
agrees with a Bethe-ansatz calculation at the particle-hole symmetric point. 
These successful comparisons validate our approach.

Let us first consider the Kondo model Eq.~\eqref{kondo}
characterized by the exchange coupling $J$ and the high-energy 
cutoff (bandwidth) $D$, with no potential scattering.
In the Wilsonian language, the RG flow corresponds to integrating the model
continuously over high-energy states thereby reducing the cutoff from
$D$ to $\Lambda$ and changing the exchange coupling constant from 
$J$ to ${\cal V}_J^{\rm Kondo} (\Lambda)$. Hence two Kondo models
are equivalent if the RG flow connects them even if their
bare values of $J$, $D$ are different. ${\cal V}_J^{\rm Kondo} (\Lambda)$
can be calculated by different means, but in the field theory language
with Abrikosov parafermions~\cite{abrikosov1965,abrikosov1970}, it is defined
from the same renormalized vertex 
as the one used in this paper and illustrated in Fig.~\ref{fig:vertex}.
The one-loop calculation gives~\cite{solyom1974,cragg1979b}
\begin{equation}\label{kondoform}
\mathcal V^{\rm Kondo}_J (\Lambda) =J -\nu_0J^2 \ln\frac{\Lambda}{D}+\ldots
\end{equation}
We already gave RG arguments in Sec.~\ref{anderkondo} showing 
that the Anderson model maps exactly onto the Kondo model for energies well below 
the charging energy $\sqrt{-\varepsilon_d(\varepsilon_d+U)}$. 
The identification between Eq.~\eqref{kondoform} and Eq.~\eqref{jver}
thus determines the coupling constants $J$ and $D$ for the Kondo model,
inherited from the Anderson model, in terms of the original parameters $t$, 
$\varepsilon_d$ and $U$.
With the arbitrary choice
\begin{equation}\label{valueD}
  D=\sqrt{-\varepsilon_d(\varepsilon_d+U)},
\end{equation}
Eq.~\eqref{kondoform} is reproduced from Eq.~\eqref{jver} if 
 $J = J_0- (\nu_0/2) J_0^2$ or
\begin{equation}\label{valueJ}
\nu_0 J= -\frac{2\Gamma \, U}{\pi \varepsilon_d(\varepsilon_d+U)}
-\frac{2 \Gamma^2 \,U^2}{\pi^2 \varepsilon_d^2(\varepsilon_d+U)^2}.
\end{equation}
The Kondo temperature $T_K$ is the energy scale for which ${\cal V}_J^{\rm Kondo} (T_K)$
is of order one. Among the different definitions of $T_K$, we use the high
temperature $T \gg T_K$ expansion~\cite{andrei1983}
of the spin susceptibility (in proper units)
\begin{equation}\label{highT}
\chi (T) = \frac{1}{4 \, T} \left( 1 - \frac{1}{\ln (T/T_K)}
- \frac{1}{2}  \frac{\ln[ \ln (T/T_K)]}{(\ln T/T_K)^2} \right).
\end{equation}
A two-loop perturbative RG calculation on the Kondo model Eq.~\eqref{kondo}
gives the Kondo temperature~\cite{hewson1997}
\begin{equation}\label{TKband}
  T_K= {\cal B} \, D \sqrt{\nu_0 \, J} \, e^{-1/\nu_0J},
\end{equation}
depending on $J$ and $D$, where ${\cal B}$ is a prefactor of order one. 
The precise value, ${\cal B}= e^{\frac34+C}/2 \pi$ where $C=0.5772\ldots$ 
is Euler's constant, is obtained by comparing
Eq.~\eqref{highT} with a weak coupling (in $J$) calculation 
of the susceptibility~\cite{andrei1981,hewson1997}. 
Substituting $J$ and $D$ in the Kondo temperature Eq.~\eqref{TKband}
by their expressions Eqs.~\eqref{valueD},~\eqref{valueJ} in terms of the 
parameters of the Anderson model, the formula Eq.~\eqref{tkhaldane}
derived by Haldane using a completely different approach~\cite{haldane1978a}
is recovered.

Sofar in this discussion, we have discarded the potential scattering term
in Eq.~\eqref{kondo} because it does not alter the RG flow
and the results of Eqs.~\eqref{valueD},~\eqref{valueJ} 
and~\eqref{tkhaldane} are still valid.
The potential scattering term can in fact be 
absorbed~\cite{cragg1978,cragg1979a,cragg1979b} into a redefinition
of the lead electron wavefunctions (with a minor negligible correction as 
 discussed in Sec.~\ref{sec:ander}) such that the standard Kondo RG flow
is recovered, albeit with electrons scattered with the phase shift
$\delta_\sigma -\pi/2 = 
- {\rm arctan} (\pi \nu_0 K(\varepsilon_d)) \simeq - \pi \nu_0 K(\varepsilon_d)$
with respect to the original electrons. 
Here $K(\varepsilon_d) =  \mathcal V^b_K$, see Eq.~\eqref{vb2}.
Using the Friedel sum rule $\sum_\sigma \delta_\sigma = \pi \, \av{\hat n}$, 
the dot occupancy reads $\av{\hat n} = 1 - 2 \nu_0 K(\varepsilon_d)$
leading, for the charge susceptibility, to the formula Eq.~\eqref{chied}
previously advertised.
In the particle-hole symmetric case, Eq.~\eqref{chied} reduces to 
\begin{equation}
  \chi_c=\frac{8\Gamma}{\pi U^2}
\left[1+\frac6\pi\left(\frac{2\Gamma}{U}\right)\right],
\end{equation}
in agreement with a Bethe-ansatz calculation~\cite{horvatic1985}.

With Eq.~\eqref{chied}, we show that the charge fluctuations
 in the Anderson model can be calculated perturbatively despite
the Kondo singularity that  affects only the spin fluctuations.
This is another indication of the spin/charge separation in the
Anderson model~\cite{tsvelick1983}. This separation no longer occurs in the presence of
a magnetic field and the problem of charge fluctuations
becomes non-perturbative. Nevertheless, the Fermi liquid approach
discussed in Sec.~\ref{sec:fermi} still applies to the
finite magnetic field case~\cite{filippone2011}.

\subsection{Field theory approach II\\ Renormalization of the Coulomb blockade model}\label{sec:matveev}

Unlike the Anderson model, the CBM does not exhibit
logarithmic singularities and the derivation of the low energy
effective model can be performed using perturbation theory.
Nevertheless, similarly to the Anderson model, an expansion
of the CBM Hamiltonian Eq.~\eqref{mat} around the zero tunneling
limit $t=0$ is unworkable because the unperturbed Hamiltonian
is not quadratic and Wick's theorem does not apply.
We thus extend the approach of Barnes by introducing one 
boson operator $b_n$ for each charge state with exactly $n$
electrons on the dot. The projection onto the physical sector
\begin{equation}
\sum_n b^\dagger_nb_n=1
\end{equation}
is realized with the chemical potential $\lambda$, as in 
Sec.~\ref{sec:anderson}, taken to infinity at the end of
calculations. In this new basis, the number of electrons
on the dot is $\hat{n} = \sum_n \, n \, b^\dagger_nb_n$ and
the charging energy part of the Hamiltonian Eq.~\eqref{mat}
reads $H_c =  E_c \sum_n \, n^2 \, b^\dagger_nb_n$.
The charge states are coupled by the tunneling term
\begin{equation}
H_T=t \sum_{kl\sigma}\left(d^\dagger_{l\sigma}c_{k\sigma} {\cal A}^\dagger
 +c^\dagger_{k\sigma}d_{l\sigma} {\cal A}   \right),
\end{equation}
where the operator ${\cal A} = \sum_n b_{n-1}^\dagger b_{n}$ removes one
electron from the dot. 

With $-E_c < \varepsilon_d < E_c$, $n=0$ defines the low energy sector
and, in a second order expansion in the dimensionless conductance
$g = N (\nu_0 t)^2$, only the charge states $n=\pm 1,\pm 2$ are
virtually occupied. We thus discard all charge states with $|n|>2$ 
from the action. The action assumes the form  $S=S_0+S_1$ with
\begin{subequations} \label{actionbos0}
\begin{align}
S_0&=\mbox{Tr}\left[-\sum_{k\sigma}c^\dagger_{k\sigma}G^{-1}_kc_{k\sigma}-\sum_{l\sigma}d^\dagger_{l\sigma}D^{-1}_ld_{l\sigma}\right],\\
S_1&=\mbox{Tr}\left[-\sum_{n=-2}
^2b^\dagger_nF^ { -1
}_nb_n+t\sum_{kl\sigma}\left(c^\dagger_{k\sigma}d_{l\sigma}\mathcal{A}+d^\dagger_{l\sigma}c_{k\sigma}\mathcal A^\dagger\right)\right],\label{actionbos}
\end{align}
\end{subequations}
where the trace stands for the typical integral over imaginary time
\begin{equation}
\mbox{Tr}\left[O\right]=\int_0^\beta d\tau O(\tau).
\end{equation}
The free propagators are given by
\begin{subequations}
  \begin{align}
  G^{-1}_k&=-\partial_\tau-\varepsilon_k,\\
  D^{-1}_l&=-\partial_\tau-\varepsilon_l,\\
  F_n^{-1}&=-\partial_\tau-\lambda-E_n,
  \end{align}
\end{subequations}
where $E_n=E_c n^2+\varepsilon_d n$ denote the bare energies of the charge
states.

\begin{figure}[htb]
  \includegraphics[width=8cm]{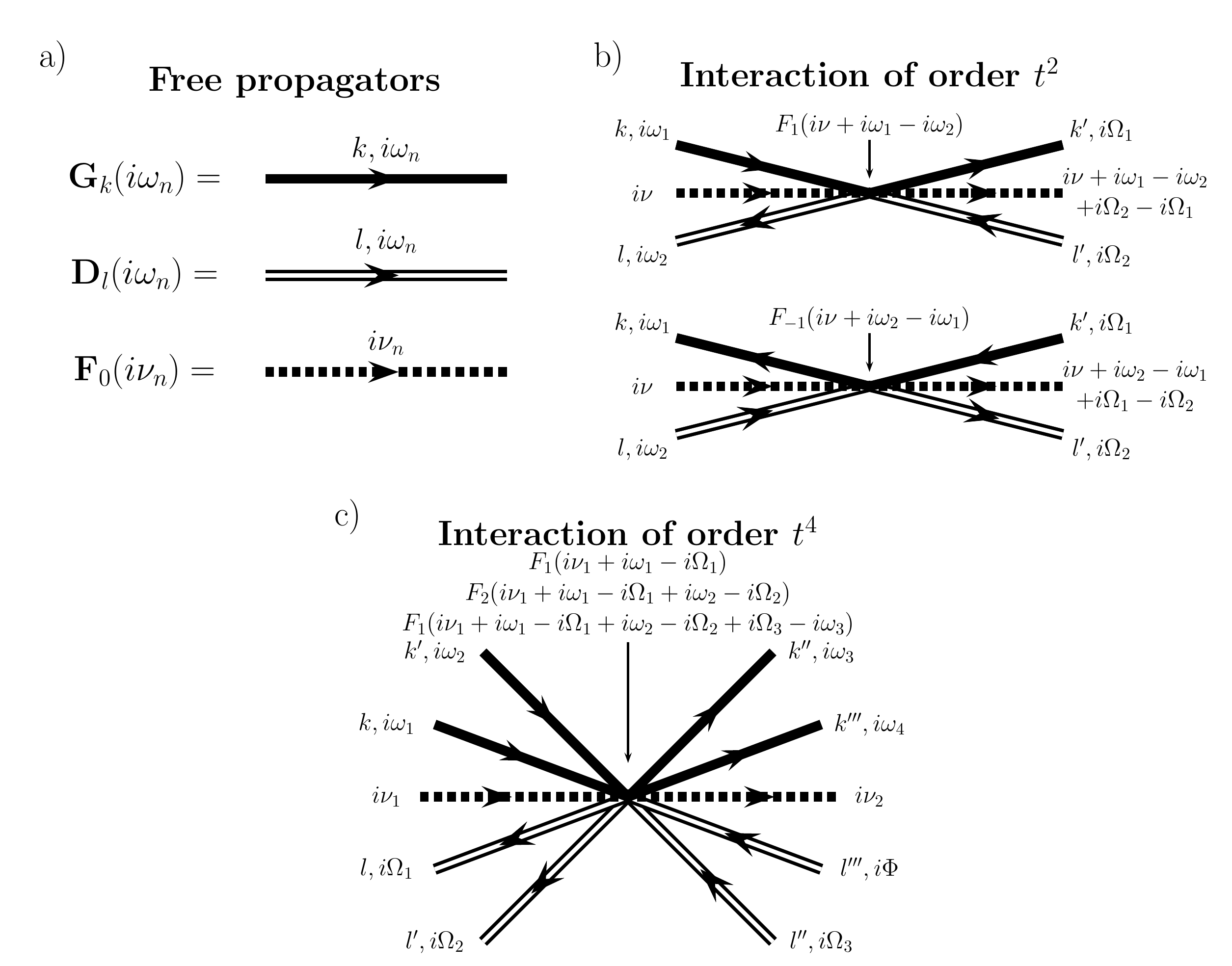}
  \caption{Diagrammatic representation of free propagators in the CBM integrated 
action (a). The diagrams are represented in the frequency
domain.  The bare interaction of order $t^2$ (b) and the one of order $t^4$ (c) are also shown.
$i\Phi=i\Omega_1-i\omega_1-i\nu_1+i\Omega_2-i\omega_2+i\omega_3-i\Omega_3+i\omega_4+i\nu_2$. 
The arrows pointing to the center of these vertices indicate the
frequency dependence of the high-energy propagators included in the
 interaction (\ref{actionmat}).}\label{fig:vertmat}
\end{figure}

The structure of the action Eq.~\eqref{actionbos0} allows 
the straightforward integration of the high-energy fields $b_{\pm1}$ and 
$b_{\pm2}$ as detailed in Appendix~\ref{appmat}. The action
then reads $S'=S_0+S_2$ where
\begin{widetext}
 \begin{equation}\label{actionmat} 
  \begin{split}
S_2=&\mbox{Tr}\left[-b^\dagger_0F^{-1}_0b_0+t^2\sum_{\stackrel{kk'll'}{\sigma\sigma'}}\left(c^\dagger_{k\sigma}d_{l\sigma}b^\dagger_0F_1d^\dagger_{
l'\sigma'}c_{k'\sigma'}b_0+d^\dagger_{l\sigma}c_{k\sigma}b^\dagger_0F_{-1}c^\dagger_{k'\sigma'}d_{
l'\sigma'}b_0\right)\right.\\
&\left.+t^4\sum_{\stackrel{kk'k''k'''ll'l''l'''}{\sigma\sigma'\sigma''\sigma'''}}c^\dagger_{k\sigma}d_{l\sigma}b^\dagger_0F_1c^\dagger_{k''\sigma''}d_{l
''\sigma''}F_2d^\dagger_{l'''\sigma'''}c_{k'''\sigma'''}F_1d^\dagger_ {
l'\sigma'}c_{k'\sigma'}b_0\right.\\
&+\left.t^4\sum_{\stackrel{kk'k''k'''ll'l''l'''}{\sigma\sigma'\sigma''\sigma'''}}d^\dagger_{l\sigma}c_{k\sigma}b^\dagger_0F_{-1}d^\dagger_
{l''\sigma''}c_{k''\sigma''}F_{-2}c^\dagger_{k'''\sigma'''}d_{l'''\sigma'''}F_{-1}c^\dagger_{k'\sigma'}d_{
l'\sigma'}b_0\right].
\end{split}
 \end{equation}
\end{widetext}
This action is the starting point of our perturbative expansion around 
$t=0$. The unperturbed action for $t=0$ is indeed quadratic and
standard diagrammatics is applicable.
 Switching to the frequency
representation of the trace,  the bare interaction of the theory 
can be drawn diagrammatically as in Fig. \ref{fig:vertmat}. 
In this section, we use the Wilsonian RG approach and introduce
the running energy scale $\Lambda$ in the decomposition $\phi = \phi^s+\phi^f$
where $\phi = b_0,c_k,d_l$. $\phi^s$ represents the slow degrees of freedom with
Matsubara frequencies $\omega_n < \Lambda$ and $\phi^f$ the fast ones, 
$\omega_ n>\Lambda$, that are integrated after expanding the action 
Eq.~\eqref{actionmat} around $t=0$. In the diagrams shown in 
Figs.~\ref{fig:selfmat},~\ref{fig:firstvertmat},~\ref{fig:sbomba1},~\ref{fig:sbomba2},
the external lines are slow modes and the internal lines fast modes.
In the standard RG treatment, the integration over fast modes is realized continuously
to follow the evolution of the coupling constants under RG. This is not
necessary in our case since no IR divergence occurs in diagrams. The fast 
modes are therefore integrated all at once and, since we are interested in
the low energy limit $\Lambda \to 0$, the energy windows for the fast
modes in fact extends over all frequencies.

\subsubsection{Low energy theory}

After averaging over the fast modes and reexponentiating the action,
one obtains the low energy form
\begin{equation}\label{lowener}
\begin{split}
 & S =S_0+\int_0^\beta d\tau\bigg[-b^\dagger_0(\tau)\mathcal F^{-1}_0b_0(\tau)+ \\[1mm]
& +\mathcal V\sum_{kk'\sigma} \left( c^\dagger_{k\sigma}(\tau)c_{k'\sigma}(\tau)
- d^\dagger_{k\sigma} (\tau) d_{k'\sigma}(\tau)  \right)
b^\dagger_0(\tau)b_0(\tau)\bigg],
\end{split}
\end{equation}
where $\mathcal F_0$ stands for the full propagator of the slave-boson $b_0$. 
The opposite sign of the lead and dot scattering terms originates from the ordering
of the operators in Eq.~\eqref{actionmat}.
The pole of the slave-boson propagator (recall that $E_0=0$)
\begin{equation}\label{propagator0}
    \mathcal F_0 (i\nu_n)= \frac{1}{i\nu_n-\lambda-\Sigma_0(i\nu_n)}
\end{equation}
defines the renormalized slave-boson energy $\tilde{E}_0 = \lambda
+ \Sigma_0 (\tilde{E}_0)$. Close to this pole, the self-energy
$\Sigma_0(i \nu_n)$ is regular and the slave-boson propagator
takes the form
\begin{equation}\label{propagator02}
    \mathcal F_\sigma(i\nu_n)= \frac{{\cal Z}_0}{i\nu_n-\tilde E_0},
\end{equation}
with ${\cal Z}_0 =[1- \partial_\omega \Sigma_0(\tilde{E}_0)]^{-1}$. Thus
the IR fixed point corresponds to   $i \nu = \tilde{E}_0$.  

The coupling constant $\mathcal V$ in the action Eq.~\eqref{lowener}
derives from the interaction vertex between slave-bosons and lead 
fermions ($-\mathcal V$ for the dot fermions), illustrated in Fig.~\ref{fig:firstvertmat}
to lowest order in $g$, taken at  $i \nu = \tilde{E}_0$ for the
slave-bosons and  $i \omega = 0$  for the fermions.
A rescaling of the slave-boson field $b_0 \to \sqrt{{\cal Z}_0} \, b_0$ 
in Eq.~\eqref{lowener}
removes ${\cal Z}_0$ from the propagator Eq.~\eqref{propagator02} 
and renormalizes the vertex $\mathcal V \to \mathcal V^R$.
The renormalized vertex $\mathcal V^R$ is therefore the relevant object
describing the scattering of electrons as in Sec.~\ref{sec:anderson}
for the Anderson model. 
After this rescaling, the limit $\lambda \to +\infty$ simply enforces
$b^\dagger_0 (\tau) b_0 (\tau)=1$ and the low energy Hamiltonian
corresponding to the action Eq.~\eqref{lowener} becomes
\begin{equation}\label{lowene2}
  H =H_0+
\mathcal V^R \sum_{kk'\sigma} \left( c^\dagger_{k\sigma} c_{k'\sigma} 
- d^\dagger_{k\sigma} d_{k'\sigma}  \right),
\end{equation}
and confirms the Fermi liquid picture developped in Sec.~\eqref{sec:large},
see for instance Eq.~\eqref{lowener0}
with $K(\varepsilon_d) = \pm \mathcal{V}_R$ for the lead/dot electrons.

Secs.~\ref{sec:slaveboson} and~\ref{sec:vertex} are 
devoted to the evaluation of $\mathcal V^R$.
The slave-boson propagator is first calculated,
in order to access ${\cal Z}_0 $ and $\tilde{E}_0$, then the vertex  $\mathcal V$.
We finally apply the Friedel sum rule to the Hamiltonian Eq.~\eqref{lowene2}
to determine the mean occupancy of the dot and compare with a direct
calculation.
%
\subsubsection{The slave-boson propagator}\label{sec:slaveboson}

\begin{figure}[h]
  \includegraphics[width=8cm]{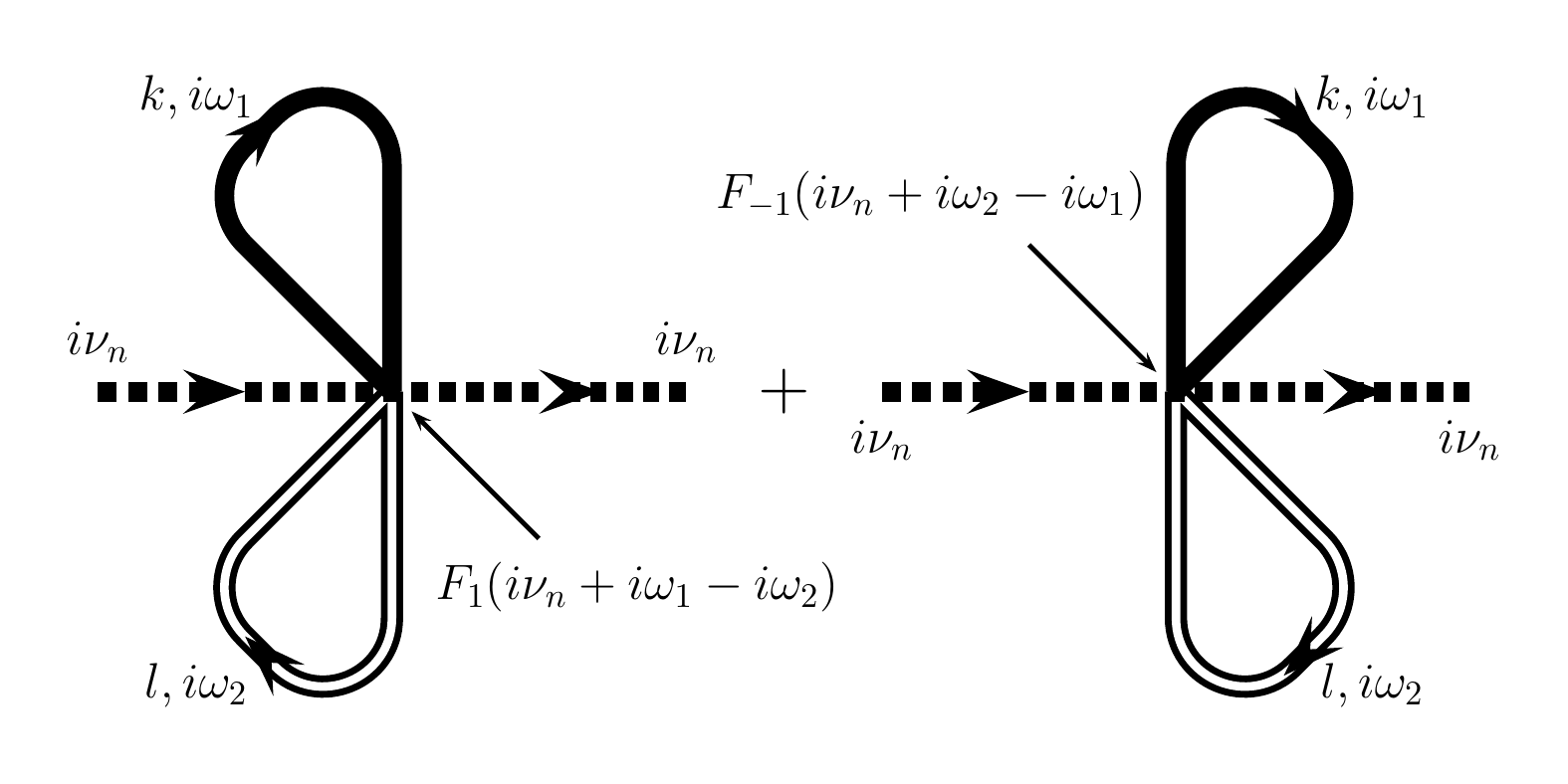}
  \caption{First order diagrams for the slave-boson self-energy.}\label{fig:selfmat}
\end{figure}

For an overall calculation of second order in $g$, only the first order
approximation of the slave-boson self-energy is needed.
Including the diagrams shown in 
Fig. \ref{fig:selfmat}, it reads
\begin{equation}\label{salf}
 \begin{split}
& \Sigma_0(i\nu_n)=-\frac{t^2}{\beta^2}\sum_{kl\sigma
i\omega_{1,2}}G_k(i\omega_1)D_l(i\omega_2)\\
&\left(F_1(i\nu_n+i\omega_1-i\omega_2)+F_{-1}(i\nu_n+i\omega_2-i\omega_1)\right)\\
&=-N(\nu_0t)^2\int d \varepsilon_1d \varepsilon_2\left[\frac{\theta(\varepsilon_1) \theta(\varepsilon_2)}{\varepsilon_1+\varepsilon_2+E_1+\lambda-i\nu_n}\right.\\
&+\left.\frac{\theta(\varepsilon_1) \theta(\varepsilon_2)}{\varepsilon_1+\varepsilon_2+E_{-1}+\lambda-i\nu_n}\right],
\end{split}
\end{equation}
where all electrons energies have been summed over and the limit
$\lambda\rightarrow\infty$ taken. The self-energy in Eq.~\eqref{salf}
exhibits a linear UV divergence. The theory is regularized with
the cutoff function $e^{-\varepsilon/D_0}$.
Fortunately, all UV divergences cancel out in the final expression of the
renormalized vertex, as discussed in Appendix~\ref{app:summa}
and the limit $D_0 \to +\infty$ is eventually taken.
From Eq.~\eqref{salf}, one extracts the values of 
${\cal Z}_0 =1 + \partial_\omega \Sigma_0(\lambda) $ and
the renormalized energy $\tilde{E}_0 = \lambda +   \Sigma_0 (\lambda)$
to order $g$.

\subsubsection{The vertex}\label{sec:vertex}
\begin{figure}[h]
  \includegraphics[width=8cm]{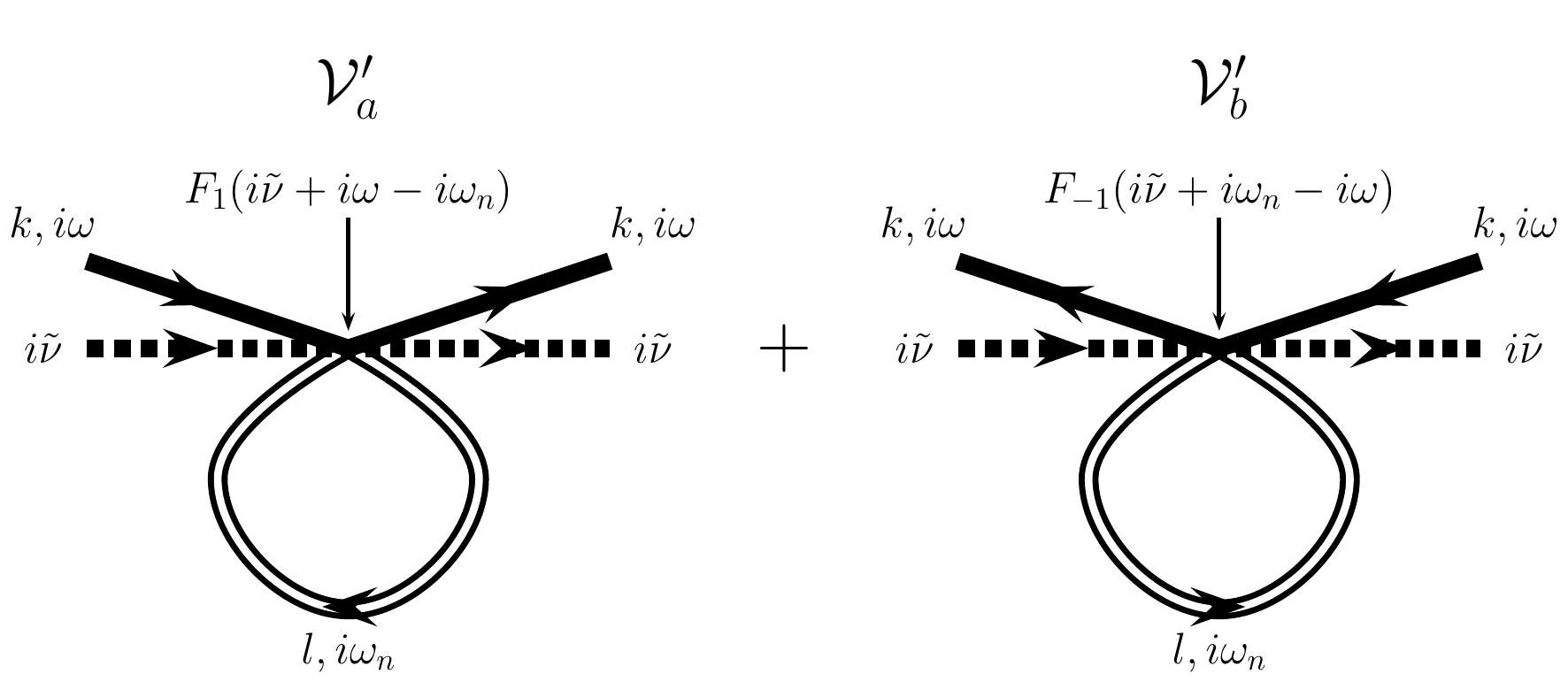}
  \caption{First order diagrams for the vertex $\mathcal V$.}\label{fig:firstvertmat}
\end{figure}

\begin{figure}[h]
  \includegraphics[width=8cm]{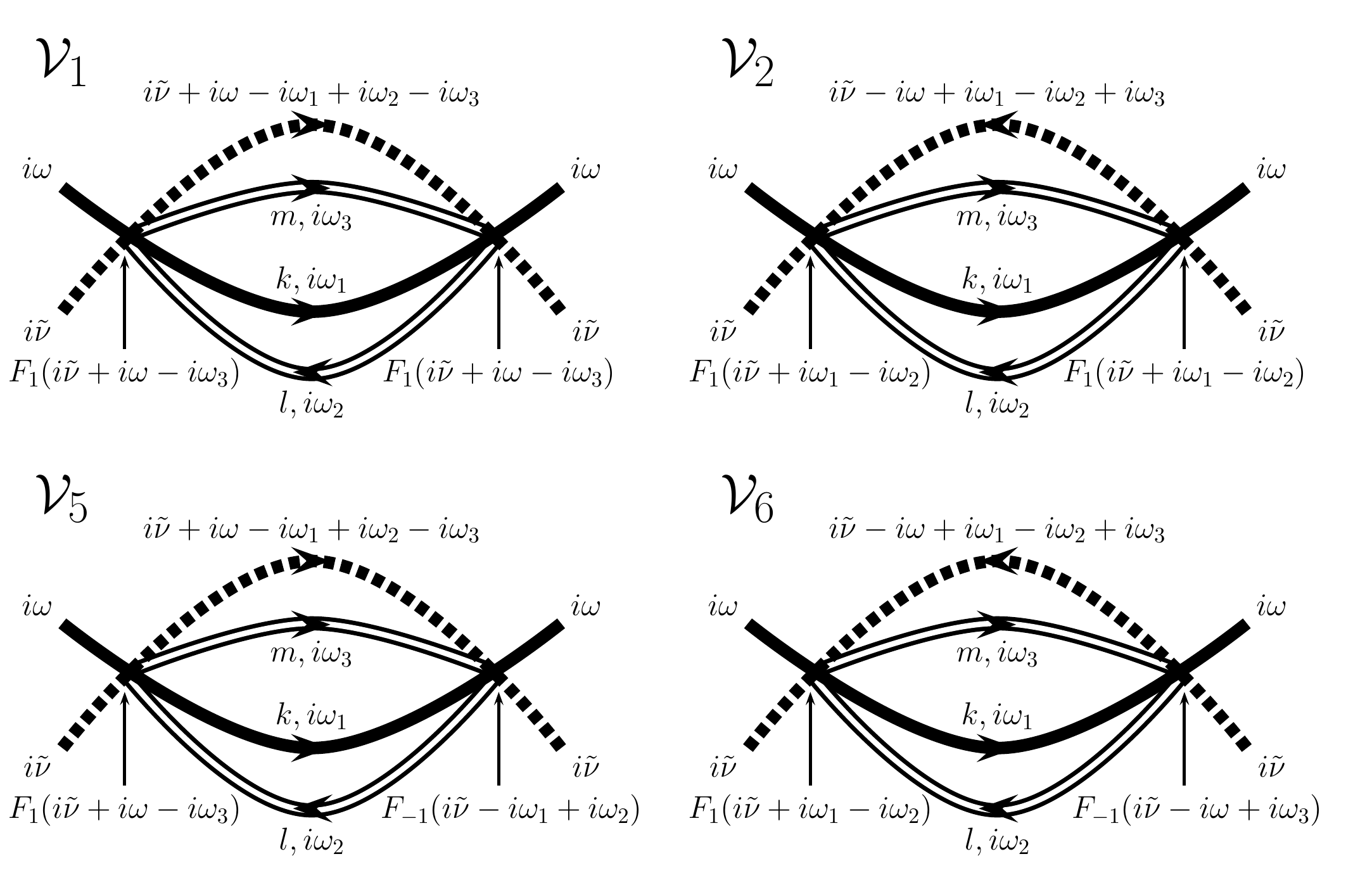}
  \caption{Second order (in $g$) diagrams involving the 
six-leg vertex shown in Fig.~\ref{fig:vertmat} and scaling
as $N t^4$. The diagrams
corresponding to the contributions $\mathcal V_3$ and $\mathcal V_4$
are not shown, they are similar to $\mathcal V_1$ and $\mathcal V_2$
but come with an opposite sign and the change $E_n\rightarrow E_{-n}$,
see also Appendix~\ref{app:summa}. The diagrams scaling only as 
$t^4$ have been discarded under the assumption of
large channel number $N$.}\label{fig:sbomba1}
\end{figure}

\begin{figure}[h]
  \includegraphics[width=8cm]{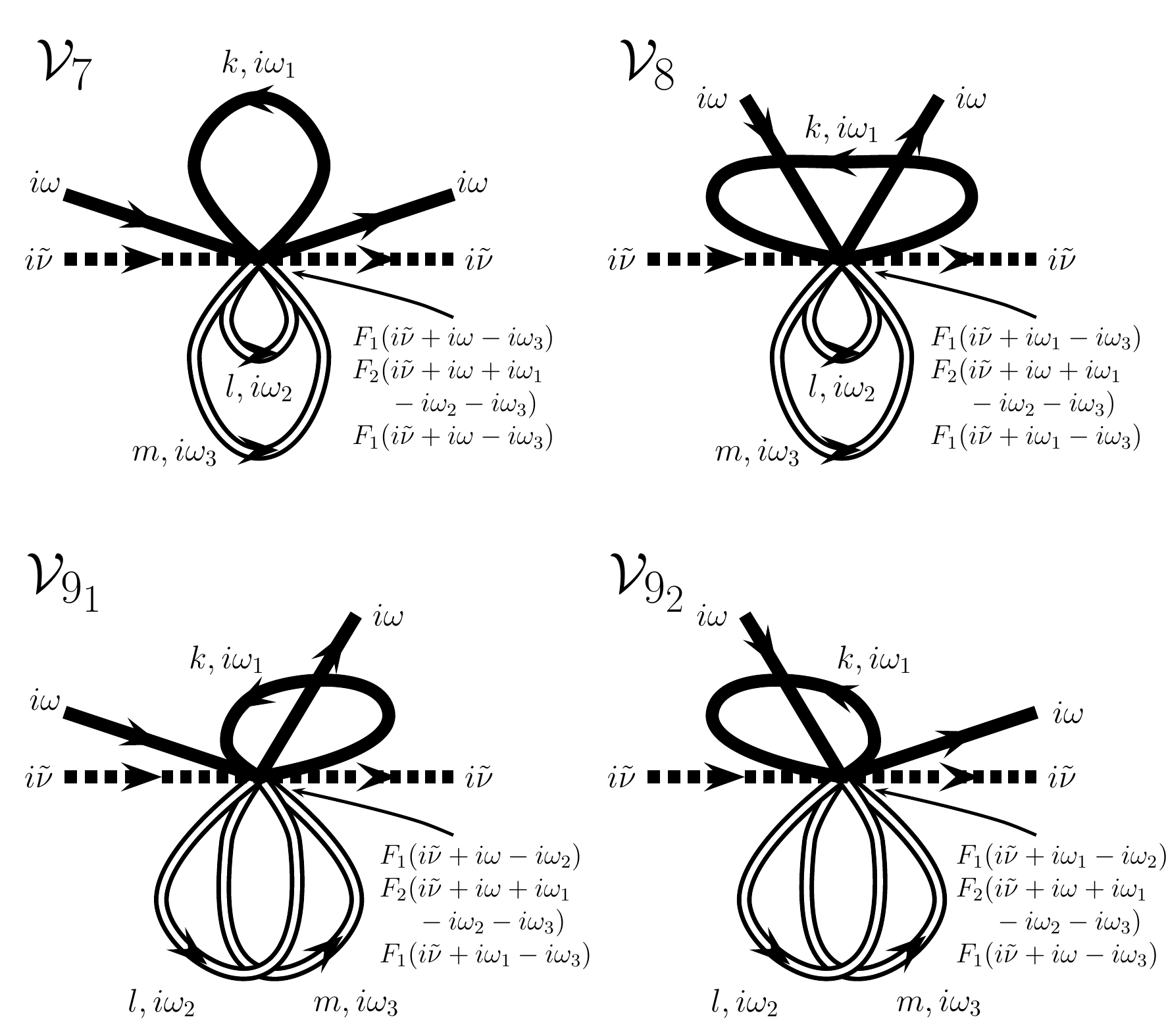}
  \caption{Same as Fig.~\ref{fig:sbomba1} but with diagrams
involving the ten-leg vertex shown in Fig.~\ref{fig:vertmat}.
The diagrams leading to $\mathcal V_{10,11,12}$ are not represented,
they are similar to $\mathcal V_{7,8,91}$
but come with an opposite sign and the change $E_n\rightarrow E_{-n}$,
see also Appendix~\ref{app:summa}. Notice $\mathcal V_9
=\mathcal V_{91}+\mathcal V_{92}=2\mathcal V_{91}$.}\label{fig:sbomba2}
\end{figure}

The leading contributions to the vertex $\mathcal V$ are represented in 
Fig.~\ref{fig:firstvertmat}. They are similar to the self-energy diagrams
of Fig.~\ref{fig:selfmat} but without contraction of the lead electron lines.
The first diagram gives
$$-\frac{t^2}{\beta}\sum_{l,i\omega_n}D_l(i\omega_n)F_1(i\nu+i\omega-i\omega_n)$$
$$ =-\nu_0t^2 \int d\varepsilon\frac{1-f(\varepsilon)}{\varepsilon+E_1-\Sigma(\lambda)},$$
\begin{equation}
 \begin{split}
&-\frac{t^2}{\beta}\sum_{l,i\omega_n}D_l(i\omega_n)F_1(i\tilde\nu+i\omega-i\omega_n)\\
 &=-\nu_0t^2 \int d\varepsilon\frac{1-f(\varepsilon)}{\varepsilon+E_1-\Sigma(\lambda)},
\end{split}
\end{equation}
where the analytical continuations $i\omega\rightarrow0$ 
and $i\nu\rightarrow\lambda+\Sigma_0 (\lambda)$ have been carried out
together with the $\lambda\rightarrow\infty$ limit.
Adding both diagrams in Fig.~\ref{fig:firstvertmat} and expanding to second
order in $g$ (fourth in $t$), we obtain 
\begin{equation}\label{giustoper}
\begin{split}
&  \mathcal V'=
\nu_0t^2\ln\frac{E_1}{E_{-1}} +N\nu_0^3t^4\int_{\varepsilon_1,\varepsilon_2,\varepsilon_3>0} 
d\varepsilon_1d\varepsilon_2d\varepsilon_3 \\[1mm]
&  \left( \sum_{s = \pm1} \frac1{\varepsilon_1+\varepsilon_2+E_s} \right)\left( \sum_{s = \pm1}
\frac{s}{(\varepsilon_3+E_{s})^2}\right).
\end{split}
\end{equation}
The number of integration variables is reduced by changing variables,
$\varepsilon_1+\varepsilon_2 \to \varepsilon_1$ and integrating over
$\varepsilon_2$, leading to
\begin{equation}\label{giustoper2}
\begin{split}
&  \mathcal V'=
\nu_0t^2\ln\frac{E_1}{E_{-1}} +N\nu_0^3t^4\int_{\varepsilon_1,\varepsilon_3>0} 
d\varepsilon_1d\varepsilon_3 \\[1mm]
&  \left( \sum_{s = \pm1} \frac{\varepsilon_1}{\varepsilon_1+E_s} \right)\left( \sum_{s = \pm1}
\frac{s}{(\varepsilon_3+E_{s})^2}\right).
\end{split}
\end{equation}
We note in passing that the leading order result, {\it i.e.} the first
term in Eq.~\eqref{giustoper2}, coincides with our previous Schrieffer-Wolff
calculation, see Eq.~\eqref{effintmat}.
An additional contribution to the renormalized vertex ${\cal V}^R$ is brought
by the slave-boson weight ${\cal Z}_0$, when the first order correction to
 ${\cal Z}_0$ multiplies the first term in Eq.~\eqref{giustoper2}, namely
\begin{equation}\label{gammaZ}
 \mathcal V''=-N\nu_0^3t^4\ln\left(\frac{E_1}{E_{-1}}\right)\,
\int_{\varepsilon>0} d \varepsilon \left( \sum_{s=\pm 1} \frac {\varepsilon}{(\varepsilon+E_s)^2}\right).
\end{equation}
We finally turn to the genuine second order diagrams for the vertex
 ${\cal V}$ and therefore ${\cal V}^R$. At this point, a distinction
can be operated between diagrams scaling as $Nt^4$ and those scaling
as $t^4$. The calculation is greatly simplified by keeping the former
and discarding the latter 
in a large $N$ calculation. In addition, the diagrams
are classified depending on whether they involve the six-leg vertex
of Eq.~\eqref{actionmat} (of order $g$) shown in Fig.~\ref{fig:vertmat},
those diagrams being listed in Fig. \ref{fig:sbomba1}, or
the ten-leg vertex (of order $g^2$), again in Fig.~\ref{fig:vertmat},
those latter diagrams being listed in Fig. \ref{fig:sbomba2}.
We shall calculate explicitly only the vertex contribution $\mathcal V_1$
and quote the results of other contributions in Appendix~\ref{app:summa}.

The expression associated to the diagram $\mathcal V_1$ is given by 
 \begin{equation} 
\begin{split}
& \mathcal V_1 =N\frac{t^4}{\beta^3}\sum_{klm,i\omega_{1,2,3}}
G_k(-i\omega_1)D_l(i\omega_2)D_m(-i\omega_3) \\[1mm]
& F_1^2(i\tilde\nu+i\omega+i\omega_3)F_0(i\tilde\nu+i\omega+i\omega_1+i\omega_2+i\omega_3),
\end{split}
 \end{equation}
and, after summing over the Matsubara frequencies $\omega_{1,2,3}$
in the limit $\lambda\rightarrow\infty$, one obtains
\begin{equation}
 \begin{split}
 \mathcal V_1&=Nt^4\sum_{klm}\frac{f(-\varepsilon_k)f(\varepsilon_l)f(-\varepsilon_m)}{(\varepsilon_m+E_1)^2(\varepsilon_l-\varepsilon_k-\varepsilon_m)}\\
 &=-N\nu_0^3t^4\int_{\varepsilon_1,\varepsilon_2>0}
 d\varepsilon_1 d \varepsilon_2 
\frac{ \varepsilon_1}{(\varepsilon_2+E_1)^2
(\varepsilon_1+\varepsilon_2)}.
 \end{split}
\end{equation}
The  calculation of the other twelve
diagrams illustrated in Figs.~\ref{fig:sbomba1}
and~\ref{fig:sbomba2} are not particularly enlightening and follow
the same line as the calculation of $\mathcal V_1$. The different
contributions are therefore summarized in Appendix~\ref{app:summa}.
The summation over these twelve terms together with $\mathcal V'$
and $\mathcal V''$ is also performed 
in Appendix~\ref{app:summa} where special attention is paid to the
cancellation of the different UV divergences.
In the limit $N\to+\infty$, the final result for
the renormalized vertex to second order in $g$ reads
 \begin{subequations}\label{A}
\begin{align}
&  \mathcal V^{R}=\nu_0t^2\ln\frac{E_1}{E_{-1}} +N\nu_0^3t^4\left(A[\varepsilon_d]-A[-\varepsilon_d]\right),\\[3mm]
  \begin{split}
& A[\varepsilon_d]=\frac{-\varepsilon_d}{2E_c}\left(\frac{4\pi^2}{3}+\ln^2\frac{E_c+\varepsilon_d}{E_c-\varepsilon_d}\right)\\[1mm] &+\frac{
8\left(2E_c^2-2E_c\varepsilon_d-\varepsilon_d^2\right) } { (3E_c+\varepsilon_d)(E_c-\varepsilon_d) } \ln\frac{E_c+\varepsilon_d}{E_c}\\[1mm]
&+\frac{(2E_c+\varepsilon_d)}{E_c}\Bigg[\ln^2\frac{E_c+\varepsilon_d}{4E_c+2\varepsilon_d}+2Li_2\left(\frac{3E_c+\varepsilon_d}{4E_c+2\varepsilon_d}\right)\\[1mm]  & -\frac{
4E_c(2E_c+\varepsilon_d)}{(E_c+\varepsilon_d)(3E_c+\varepsilon_d)}\ln\frac{4E_c+2\varepsilon_d}{E_c}\Bigg].
\end{split}
\end{align}
 \end{subequations}
This result, substituted in the low energy Hamiltonian Eq.~\eqref{lowene2},
gives access to the dot occupancy 
\begin{equation}\label{occupancy}
\av {\hat n} = g\ln\frac{E_c-\varepsilon_d}{E_c+\varepsilon_d}-g^2\left(A[\varepsilon_d]-A[-\varepsilon_d]\right),
\end{equation}
by using the Friedel sum rule $$\av {\hat n} = - (N/\pi) \,
{\rm arctan} (\pi \nu_0 \mathcal V^{R} ) \simeq - N \nu_0 \mathcal V^{R}.$$
 The result of Eq.~\eqref{occupancy} coincides with a direct
 calculation of the dot occupancy~\cite{grabert1994a,grabert1994b} 
using a different perturbative approach, which validates, at least
perturbatively, the Fermi liquid description emphasized 
in Sec.~\ref{sec:fermi}.

\section{Summary and conclusions}

We investigated the dynamical response of a quantum dot attached to a lead. 
We studied the low frequency charge fluctuations on the dot 
that are related to the admittance of this quantum circuit in the linear regime.
We argued that the system at low energy behaves as a local Fermi liquid
where inelastic scattering events can be disregarded and lead electrons 
are simply
coherently backscattered by the dot. The present work extends in a way 
the general analysis built by B\"uttiker and coworkers to study the dynamic
admittance of mesoscopic conductors, by including arbitrarily strong Coulomb
interactions within the dot. We avoid the introduction of an approximate 
self-consistent potential on the dot.

By computing the power dissipated by the external AC drive, we were able to derive
a set of general formulas for the quantum capacitance and the charge relaxation
resistance that characterize the response of the dot at low frequency. Remarkably,
the results are essentially not so different from the weakly interacting picture.
The fundamental reason is that electrons close to the Fermi energy do not feel
strong interactions as a result of Pauli blocking, following the standard
Fermi liquid argument~\cite{luttinger1961}. Our approach is naturally not applicable
for models exhibiting non-Fermi liquid physics such as the two-channel
model close to charge degeneracy. In these models, the mere definition 
of a capacitance and a charge relaxation resistance is elusive as a result
of unconventional scaling laws.

Another important assumption in our work
is that the Friedel sum rule is satisfied. This seems to be the case for
the two models we investigated, but it is certainly not general,
even for a Fermi liquid fixed point. However, the fact that dissipation comes
from the time-dependence in the phase shift felt by lead electrons does not
rely on the Friedel sum rule and should apply to more general models such as
double-dot geometries. Also we only considered the case of zero temperature
or temperatures much smaller than the charging energy. At higher temperatures,
 inelastic modes are excited and the analysis presented here is not applicable.


In order to put the Fermi liquid picture on firm grounds, we calculated, based
on a renormalization-group analysis, the low energy
theory for two representative models describing the quantum RC circuit: the 
Anderson model and the Coulomb blockade model. In addition to providing
an alternative demonstration of the mapping from the Anderson model 
to the Kondo model,
we find that the phenomenological low energy theory proposed in this 
paper is recovered
perturbatively for the two models. The Friedel sum rule was also 
checked explicitly.

We conclude with a technical remark regarding Ref.~\cite{rodionov2009}
where the case of a tunnel junction (infinite $N$) was considered.
The definition of the charge relaxation resistance $R_q$ involves
formally the limit of vanishing frequency $\omega \to 0$.
At finite temperature, this limit does not commute with 
an expansion in the tunneling $g$ or $\Gamma$ as emphasized
in Ref.~\cite{rodionov2009} (see also Ref.~\cite{gotze1971}).
At zero temperature however, the two limits commute~\footnote{
this can be seen, for example, by noting that the
singular term  $\propto 1/\omega$ given by Eq.($75$) in Ref.~\cite{rodionov2009}
vanishes at zero temperature.}
 and the
perturbative expansion of Sec.~\ref{sec:renormalization}
becomes justified.
Moreover, at finite by 
small temperatures, a large frequency $\hbar \omega \gg k_B T$ is
sufficient to suppress the singular term  $\propto 1/\omega$
and the results of Sec.~\ref{sec:renormalization} are still valid.

We thank K. Le Hur and L. Glazman for stimulating discussions.

\appendix

\begin{widetext}
\section{Calculation of the vertex $\mathcal V^a$}\label{app:vertex}
Once considered
\begin{equation}
\left(\mathbf S_{\beta\tau}\cdot\mathbf s_{\alpha\sigma}\right)\left(\mathbf S_{\tau'\beta}\cdot\mathbf s_{\sigma'\alpha}\right)=-\frac12\mathbf
S_{\tau'\tau}\cdot s_{\sigma'\sigma}+\frac3{16}\delta_{\tau\tau'}\delta_{\sigma\sigma'},
\end{equation}
it is possible to determine which term will contribute to the exchange and to the potential scattering part of the vertex $\mathcal V^a=\mathbf
S\cdot\mathbf s \mathcal V^a_J+\mathcal V^a_K$, where
\begin{subequations}
  \begin{align}
\mathcal V^a_J&=4\frac{t^4}\beta\sum_{k,i\omega_n}F(i\omega_n)G_k(i\omega_\Lambda+i\omega-i\omega_n)\left[
F_2^2(i\omega_\Lambda+i\omega)+F_2(i\omega_\Lambda+i\omega)F_0(i\omega_n-i\omega)\right],\\
\mathcal V^a_K&=-\frac{t^4}\beta\sum_{k,i\omega_n}F(i\omega_n)G_k(i\omega_\Lambda+i\omega-i\omega_n)\left[
F_0^2(i\omega_n-i\omega)+F_2^2(i\omega_\Lambda+i\omega)+F_0(i\omega_n-i\omega)F_2(i\omega_\Lambda+i\omega)\right].
 \end{align}
\end{subequations}
We detail the calculation of the Matsubara sums only for the following example:
\begin{multline}
 \frac1\beta\sum_{k,i\omega_n}F(i\omega_n)G_k(i\omega_\Lambda+i\omega-i\omega_n)F_0^2(i\omega_n-i\omega)=\nu_0\int_{-D_0}^{D_0}d\varepsilon\left[\frac{
f(\lambda)}{(i\omega_\Lambda+i\omega-\lambda-\varepsilon)(\varepsilon_d-i\omega)^2}-\right.\\
 \left.-\frac{f(i\omega_\Lambda+i\omega-\varepsilon)}{(i\omega_\Lambda+\varepsilon_d-\lambda-\varepsilon)^2(i\omega_\Lambda+i\omega-\varepsilon-\lambda)}+\left.\frac{d}{dz}\frac{
f(z)}{(z-\lambda)(i\omega_\Lambda+i\omega-z-\varepsilon)}\right|_{z=i\omega+\lambda-\varepsilon_d}\right].
\end{multline}
Notice $f(i\omega_\Lambda+i\omega-\varepsilon)=f(-\varepsilon)$. The analytical continuations of Eqs. (\ref{ancontinu}) can now be performed and the first
deviation from $\lambda$ of $\tilde \varepsilon_d$ in Eq. (\ref{pole}) can be neglected to this order. The denominators do not depend
on $\lambda$ anymore and, as required by the projection technique, the $\lambda\rightarrow\infty$ limit can be taken. All the contributions of the poles
which were proportional to $\lambda$ disappear and the only remaining integral is
$$\nu_0\int_{0}^{D_0}\frac{d\varepsilon}{(\varepsilon+\Lambda)(\varepsilon_d-\varepsilon)^2}=\frac{\nu_0}{\varepsilon_d^2}\left(\ln\frac{-\varepsilon_d}{D_0}-\ln\frac{
\Lambda}
{D_0}-1\right).
$$
It is also possible to take the infinite bandwidth limit $D_0\rightarrow\infty$ in this expression, but it is convenient to keep a finite cutoff for the moment. The same kind of calculations gives analog results for the remaining sums
\begin{eqnarray}
\frac1\beta\sum_{k,i\omega_n}F(i\omega_n)G_k(i\omega_\Lambda+i\omega-i\omega_n)F_0(i\omega_n-i\omega)F_2(i\omega_\Lambda+i\omega)&=&\frac{\nu_0}{
\varepsilon_d(\varepsilon_d+U)}\left(\ln\frac{\Lambda}{D_0}-\ln\frac{-\varepsilon_d}{D_0}\right),\\ 
\frac1\beta\sum_{k,i\omega_n}F(i\omega_n)G_k(i\omega_\Lambda+i\omega_i\omega_n)F^2_2(i\omega_\Lambda+i\omega)&=&-\frac{\nu_0}{(\varepsilon_d+U)^2}\ln\frac{\Lambda}{D_0}.
\end{eqnarray}
These last expressions allow us to obtain Eqs. (\ref{va}).
\end{widetext}

\section{Integration of the high energy modes in the Coulomb blockade model}\label{appmat}
The action Eq. (\ref{actionmat}) being quadratic in the high 
energy modes $b_{\pm2}$, their integration is straightforward and 
brings separate terms to the effective action
\begin{subequations}
  \begin{align} &b_2:t^2\sum_{kk'll'\sigma\sigma'}\mbox{Tr}\left[c^\dagger_{k\sigma}d_{l\sigma}b^\dagger_1F_2d^\dagger_{l'\sigma'}c_{k'\sigma'}b_1\right],\\
 &b_{-2}:t^2\sum_{kk'll'\sigma\sigma'}\mbox{Tr}\left[d^\dagger_{l\sigma}c_{k\sigma}b^\dagger_{-1}F_{-2}c^\dagger_{k'\sigma'}d_{l'\sigma'}b_{-1}\right].
 \end{align}
\end{subequations}
This action can be written as $S=S_0+S_1+S_{-1}$, with
\begin{subequations}
 \begin{align}
 \begin{split}
S_{1}=\mbox{Tr}&\left[-b^\dagger_1\Phi^{-1}_1
b_1\right.\\&\quad\left.+t\sum_{kl\sigma}\left(c^\dagger_{k\sigma}d_{l\sigma}b^\dagger_0b_1+d^\dagger_{l\sigma}c_{k\sigma}b^\dagger_1 b_0\right)\right],
 \end{split}\\
\begin{split}
 S_{-1}=\mbox{Tr}&\left[-b^\dagger_{-1}\Phi^{-1}_{-1}
b_{-1}\right.\\&\left.+t\sum_{kl\sigma}\left(c^\dagger_{k\sigma}d_{l\sigma}b^\dagger_{-1}b_0+d^\dagger_{l\sigma}c_{k\sigma}b^\dagger_{0} b_{-1}\right)\right],
\end{split}
\end{align}
\end{subequations}
and the new effective propagators
\begin{eqnarray}
 \Phi^{-1}_{1}&=&F^{-1}_1-t^2\sum_{kk'll'\sigma\sigma'}c^\dagger_{k\sigma}d_{l\sigma}F_2d^\dagger_{l'\sigma'}c_{k'\sigma'},\\
 \Phi^{-1}_{-1}&=&F^{-1}_{-1}-t^2\sum_{kk'll'\sigma\sigma'}d^\dagger_{l\sigma}c_{k\sigma}F_{-2}c^\dagger_{k'\sigma'}d_{l'\sigma'}.
\end{eqnarray}
The integration over the $b_{\pm1}$ modes is still Gaussian and the 
effective action without high energy bosonic modes is finally obtained
\begin{equation}
\begin{split}
 S =& S_0+\mbox{Tr}\left[\right.-b^\dagger_0F^{-1}_0b_0+\\[2mm]
& t^2\sum_{kk'll'\sigma\sigma'}c^\dagger_{k\sigma}d_{l\sigma}b^\dagger_0\Phi_1d^\dagger_{l'\sigma'}c_{k'\sigma'}b_0+\\
&  \left.t^2\sum_{kk'll'\sigma\sigma'}d^\dagger_{l\sigma}c_{k\sigma}b^\dagger_0\Phi_{-1}c^\dagger_{k'\sigma'}d_{l'\sigma'}b_0\right].
\end{split}
\end{equation}
The $\Phi$ operators can be expanded perturbatively in $t$
\begin{subequations}
 \begin{align}
   \Phi_1&=F_1+t^2\sum_{kk'll'\sigma\sigma'}F_1c^\dagger_{k\sigma}d_{l\sigma}F_2d^\dagger_{l'\sigma'}c_{k'\sigma'}F_1,\\
  \Phi_{-1}&=F_{-1}+t^2\sum_{kk'll'\sigma\sigma'}F_{-1}d^\dagger_{l\sigma}c_{k\sigma}F_{-2}c^\dagger_{k'\sigma'}d_{l'\sigma'}F_{-1}.
  \end{align}
\end{subequations}
This gives the action Eq. (\ref{actionmat}).

\section{Summation of all the contributions of $\mathcal V^R$ in the Coulomb blockade model}\label{app:summa}
All the contributions corresponding to the diagrams of Figs. \ref{fig:sbomba1} and \ref{fig:sbomba2} are listed below (with ${\cal C} = N\nu_0^3t^4$)
\begin{eqnarray}\label{listed}
  \mathcal V_1&=&- {\cal C} \int d\varepsilon_1 d\varepsilon_2\frac {\varepsilon_1}{(\varepsilon_2+E_1)^2(\varepsilon_1+\varepsilon_2)},\nonumber\\
  \mathcal V_2&=&{\cal C}\int d\varepsilon_1d\varepsilon_2\frac {\varepsilon_1}{(\varepsilon_1+E_1)^2(\varepsilon_1+\varepsilon_2)},\nonumber\\
  \mathcal V_3&=&{\cal C}\int d\varepsilon_1d\varepsilon_2\frac {\varepsilon_1}{(\varepsilon_2+E_{-1})^2(\varepsilon_1+\varepsilon_2)},\nonumber\\
  \mathcal V_4&=&-{\cal C}\int d\varepsilon_1d\varepsilon_2\frac{\varepsilon_1}{(\varepsilon_1+E_{-1})^2(\varepsilon_1+\varepsilon_2)},\nonumber\\
  \mathcal V_5&=&-2{\cal C}\int d\varepsilon_1d\varepsilon_2 \frac {\varepsilon_1}{(\varepsilon_1+\varepsilon_2)(\varepsilon_2+E_1)(\varepsilon_1+E_{-1})},\nonumber\\
  \mathcal V_6&=&2{\cal C}\int d\varepsilon_1d\varepsilon_2\frac {\varepsilon_1}{(\varepsilon_1+\varepsilon_2)(\varepsilon_1+E_1)(\varepsilon_2+E_{-1})},\nonumber\\
  \mathcal V_7&=&-{\cal C}\int d\varepsilon_1d\varepsilon_2 \frac {\varepsilon_1}{(\varepsilon_2+E_1)^2(\varepsilon_1+\varepsilon_2+E_2)},\nonumber\\
  \mathcal V_8&=&-{\cal C}\int d\varepsilon_1d\varepsilon_2 \frac {\varepsilon_1}{(\varepsilon_1+E_1)^2(\varepsilon_1+\varepsilon_2+E_2)},\nonumber\\
  \mathcal V_9&=&-2{\cal C}\int d\varepsilon_1d\varepsilon_2 \frac {\varepsilon_1}{(\varepsilon_2+E_1)(\varepsilon_1+\varepsilon_2+E_2)(\varepsilon_1+E_1)},\nonumber\\
  \mathcal V_{10}&=&{\cal C}\int d\varepsilon_1d\varepsilon_2 \frac {\varepsilon_1}{(\varepsilon_2+E_{-1})^2(\varepsilon_1+\varepsilon_2+E_{-2})},\nonumber\\
  \mathcal V_{11}&=&{\cal C}\int d\varepsilon_1d\varepsilon_2 \frac {\varepsilon_1}{(\varepsilon_1+E_{-1})^2(\varepsilon_1+\varepsilon_2+E_{-2})},\nonumber\\
  \mathcal V_{12}&=&2{\cal C}\int d\varepsilon_1d\varepsilon_2 \frac {\varepsilon_1}{(\varepsilon_1+E_{-1})(\varepsilon_1+\varepsilon_2+E_{-2})(\varepsilon_2+E_{-1})},\nonumber\\
\end{eqnarray}
where integrals run over the $\varepsilon_{1,2}>0$ domain,
to which the contributions of Eq.~\eqref{giustoper2} and Eq.~\eqref{gammaZ}
must be added. Whereas each term in
Eqs.~\eqref{listed},~\eqref{giustoper2} and~\eqref{gammaZ}
suffers from an UV divergence, the summation over all contributions
is finite and does not depend on the cutoff procedure.
 We shall adopt a sharp cutoff  
at energy $D_0$ in the following.
Moreover, the calculation exhibits a particle-hole symmetry: 
for example, $\mathcal V_3$ can be viewed as the
symmetric of $\mathcal V_1$, they have opposite sign and 
$E_n$ exchanged with $E_{-n}$. The result will then be necessarily 
of the form $A[\varepsilon_d]-A[-\varepsilon_d]$, which implies 
that any constant independent of $\varepsilon_d$
will be ignored during calculations. The dilogarithm function appears 
\begin{equation}
 Li_2(z)=\int_z^0dt\frac{\ln(1-t)}t, 
\end{equation}
and the following equalities will be exploited
\begin{subequations}
 \begin{align}
 &Li_2(x)+Li_2(1-x)=\frac{\pi^2}6-\ln x\ln(1-x),\\
 &Li_2(x)+Li_2\left(\frac1x\right)=\frac{\pi^2}3-\frac12\ln^2x-i\pi\ln x, ~~(x\geq 1).
 \end{align}
\end{subequations}
As an intermediate step, we find (we omit the $N\nu_0^3t^4$ factor)
\begin{align*}
  \begin{split}
  \mathcal V_1+\mathcal V_2&=-\frac {D_0}{E_1}-2\ln E_1\ln D_0+\ln^2D_0\\&\quad-2\ln D_0+\ln^2E_1+2\ln E_1,
  \end{split}\\
  \begin{split}
  \mathcal V_7+\mathcal V_8&=-\frac {D_0}{E_1}+\frac {E_2}{E_1}\ln D_0+\\&\quad\frac1{E_1(E_2-E_1)}(E_1E_2\ln E_1-E_2^2\ln E_2),
  \end{split}\\
  \begin{split}
  \mathcal V_5+\mathcal V_6+\mathcal V''&=-\frac{\varepsilon_d}{E_c}\pi^2-\frac{\varepsilon_d}{E_c}\ln^2\frac {E_1}{E_{-1}}+2\ln \frac{E_1}{E_{-1}},
  \end{split}\\
  \begin{split}
  \mathcal V_9&=2\ln E_1\ln D_0-\ln^2D_0+\frac{E_1}{E_c}\ln^2E_1\\&\quad-\frac{E_1}{E_c}\frac{\pi^2}2+\frac{E_2}{E_c}\frac{\pi^2}6+\frac{E_2}{E_c}Li_2\left(\frac{E_2-E_1}{E_2}\right)\\&\quad+\frac
{E_2}{2E_c}\ln^2E_2-\frac{E_2}{E_c}\ln E_1\ln E_2,
\end{split}
\end{align*}
with the contribution of Eq. (\ref{giustoper2})
\begin{align}
   \mathcal V'&=\nu_0t^2\ln\frac{E_1}{E_{-1}}+N\nu_0^3t^4\left(\mathcal V'_a+\mathcal V'_b\right),\\
    \begin{split}
  \mathcal V'_a&=\int d\varepsilon_1d\varepsilon_2\frac{1}{(\varepsilon_2+E_1)}\left(\frac{\varepsilon_1}{(\varepsilon_1+E_1)}+\frac{\varepsilon_1}{(\varepsilon_1+E_{-1})}\right)\\
  &=\frac{2D_0}{E_1}-\ln D_0-\frac {E_{-1}}{E_1}\ln D_0+\ln E_1+\frac {E_{-1}}{E_1}\ln E_{-1},
  \end{split}
\end{align}
where
$\mathcal V'_b$ is obtained from $\mathcal V_a'$ by particle-hole symmetry. 
It can be checked explicitly that the terms depending on the
cutoff $D_0$ in the above expressions cancel out when the summation over all
contributions is carried out. One is left with
\begin{align*}
  \mathcal V_1+\mathcal V_2&=\ln^2E_1+2\ln E_1,\\
  \mathcal V_7+\mathcal V_8&=\frac1{E_1(E_2-E_1)}(E_1E_2\ln E_1-E_2^2\ln E_2),\\
  \mathcal V'_a&=\ln E_1+\frac {E_{-1}}{E_1}\ln E_{-1},\\
  \mathcal V_5+\mathcal V_6+\mathcal V''&=-\frac{\varepsilon_d}{E_c}\pi^2-\frac{\varepsilon_d}{E_c}\ln^2\frac {E_1}{E_{-1}}+2\ln\frac{E_1}{E_{-1}},\\
  \mathcal V_9&=\frac{E_1}{E_c}\ln^2E_1-\frac{E_1}{E_c}\frac{\pi^2}2+\frac{E_2}{E_c}\frac{\pi^2}6+\frac
{E_2}{2E_c}\ln^2E_2\\&\quad+\frac{E_2}{E_c}Li_2\left(\frac{E_2-E_1}{E_2}\right)-\frac{E_2}{E_c}\ln E_1\ln E_2.\\ 
\end{align*}
Adding the particle-hole symmetric terms, one finally arrives at Eq.~\eqref{A}.

\bibliographystyle{apsrev4-1}
\bibliography{bibliographie}

\end{document}